\documentclass[useAMS,usenatbib]{mn2e}
\usepackage{mn2e-breakabs}
\usepackage{graphicx}
\usepackage{times}
\def\la{\mathrel{\mathpalette\fun <}}

\def\fun#1#2{\lower3.6pt\vbox{\baselineskip0pt\lineskip.9pt
        \ialign{$\mathsurround=0pt#1\hfill##\hfil$\crcr#2\crcr\sim\crcr}}}
        
\def\rmFoM{{\rm FoM}}

\def\bfp{\mbox{\bf p}}
\def\bfq{\mbox{\bf q}}
\def\bfk{\mbox{\bf k}}
\def\bfr{\mbox{\bf r}}
\def\etal{{\frenchspacing\it et al.}}

\def\rmdet{{\rm det}}
\def\rmCov{{\rm Cov}}
\def\rmFoM{{\rm FoM}}

\def\rd{{\rm d}}

\newcommand{\be}{\begin{equation}}
\newcommand{\ee}{\end{equation}}
\newcommand{\ba}{\begin{eqnarray}}
\newcommand{\ea}{\end{eqnarray}}
\newcommand{\simgt}{\,\hbox{\lower0.6ex\hbox{$\sim$}\llap{\raise0.6ex\hbox{$>$}}}\,}
\newcommand{\simlt}{\,\hbox{\lower0.6ex\hbox{$\sim$}\llap{\raise0.6ex\hbox{$<$}}}\,}

\begin{document}

\title[Designing a Galaxy Redshift Survey]
{Designing a space-based galaxy redshift survey \\
to probe dark energy}

\author[Yun Wang et al.]{
  \parbox{\textwidth}{
    Yun Wang$^{1}$\thanks{E-mail: wang@nhn.ou.edu},
 Will~Percival$^2$, 
 Andrea~Cimatti$^3$, 
 Pia~Mukherjee$^4$, 
Luigi~Guzzo$^5$, 
Carlton M. Baugh$^6$, 
Carmelita Carbone$^3$, 
Paolo Franzetti$^7$, 
Bianca Garilli$^7$, 
James E. Geach$^6$, 
Cedric G. Lacey$^6$,
Elisabetta Majerotto$^5$, 
Alvaro Orsi$^6$, 
Piero Rosati$^8$, 
Lado Samushia$^{2,9}$, 
Giovanni~Zamorani$^{10}$
  }
  \vspace*{4pt} \\
  $^1$Homer L. Dodge Department of Physics \& Astronomy, Univ. of Oklahoma,
                 440 W Brooks St., Norman, OK 73019, U.S.A.\\
$^2$Institute of Cosmology and Gravitation, University of                      
Portsmouth, Dennis Sciama building, Portsmouth, P01 3FX, U.K.\\
$^3$Dipartimento di Astronomia, Alma Mater Studiorum - Universit\a`a di Bologna,                     
Via Ranzani 1, I-40127, Italy\\
$^4$Sussex Astronomy Centre, Dept. of Physics \& Astronomy, University of Sussex,
Falmer, Brighton, BN1 9QH, U.K.\\
$^5$INAF - Osservatorio di Brera, Via Bianchi 46, I-23807 Merate, Italy\\
$^6$Institute for Computational Cosmology, Physics Department, Durham University, South Road,
Durham DH1 3LE, U.K.\\
$^7$INAF, IASF-Milano, via Bassini 15, 20133 Milano, Italy \\
$^8$European Southern Observatory, Karl Schwarzschild Strasse 2, Garching bei       
Muenchen, D-85748, Germany\\
$^9$National Abastumani Astrophysical Observatory, Ilia State               
University, 2A Kazbegi Ave, GE-0160 Tbilisi, Georgia\\
$^{10}$ INAF - Osservatorio Astronomico di Bologna, 
via Ranzani 1, I-40127 Bologna, Italy}

\date{\today} 

\maketitle

\begin{abstract}

  A space-based galaxy redshift survey would have enormous power in
  constraining dark energy and testing general relativity, provided
  that its parameters are suitably optimized.  We study viable
  space-based galaxy redshift surveys, exploring the dependence of the
  Dark Energy Task Force (DETF) figure-of-merit (FoM) on redshift
  accuracy, redshift range, survey area, target selection, and
  forecast method. Fitting formulae are provided for convenience. We
  also consider the dependence on the information used: the full
  galaxy power spectrum $P(k)$, $P(k)$ marginalized over its shape, or
  just the Baryon Acoustic Oscillations (BAO).  We find that the
  inclusion of growth rate information (extracted using redshift space
  distortion and galaxy clustering amplitude measurements) leads to a
  factor of $\sim$ 3 improvement in the FoM, assuming general
  relativity is not modified. This inclusion partially compensates for
  the loss of information when only the BAO are used to give geometrical 
  constraints, rather than using the full $P(k)$ as a standard ruler. 
  We find that a space-based galaxy redshift survey
  covering $\sim$20,000\,deg$^2$ over $0.5\la z\la 2$ with
  $\sigma_z/(1+z)\leq 0.001$ exploits a redshift range that is only easily
  accessible from space, extends to sufficiently low redshifts to
  allow both a vast 3-D map of the universe using a single tracer
  population, and overlaps with ground-based surveys to enable robust
  modeling of systematic effects.  We argue that these parameters are
  close to their optimal values given current instrumental and
  practical constraints.

\end{abstract}

\begin{keywords}
  cosmology: observations, distance scale, large-scale structure of
  Universe
\end{keywords}

\section{Introduction}  \label{sec:intro}

More than a decade after the discovery of cosmic acceleration \citep{Riess98,Perl99},
its cause (dubbed ``dark energy'' for convenience) remains shrouded in mystery. 
While current observational data are consistent with dark energy being a
cosmological constant, the uncertainties are large, and do not rule
out models with dynamical scalar fields 
(see, e.g., \citealt{Freese87,Linde87,Peebles88,Wett88,Frieman95,Caldwell98,Kaloper06,Chiba09}),
or models that modify general relativity (see e.g., 
\citealt{SH98,Parker99,Boisseau00,DGP00,Freese02,Capozziello05,Pad08,Kahya09,OCallaghan09}). 
For recent reviews, see 
\cite{Maartens04,Copeland06,Ruiz07,Ratra07,Frieman08,Caldwell09,Uzan09,Woodard09,Wang10}.
Several ground-based and space-born experiments have been proposed to
determine the nature of cosmic acceleration through tight control of systematic effects
and high statistical precision using multiple techniques.

A galaxy redshift survey in the near-IR from space provides a powerful probe of
dark energy and gravity, and has four key advantages over ground based surveys:
\begin{enumerate}
\item the ability to easily measure redshifts for galaxies
at $z>1$, especially in the so called ``redshift desert'' at $1.3<z<2$,
given the low near-IR background, 
\item the ability to measure redshifts for in both hemispheres in a
  single survey,
\item homogeneous dataset and low-level of systematics
such as seeing and weather induced fluctuations in efficiency,
\item the speed of the survey (e.g. about                             
4 years to cover 20,000 deg$^2$, see e.g. Laureijs et al. 2009). 
  
\end{enumerate}
Two proposed dark energy space missions, 
Euclid\footnote{http://sci.esa.int/euclid}
and JDEM\footnote{http://jdem.gsfc.nasa.gov/}, 
are being considered by ESA and NASA/DOE respectively.

Two main approaches have been considered so far for space-based                               
massive spectroscopic surveys. The first is to use ``multi-slit''                               
spectroscopy aimed at observing a pure magnitude-limited sample of galaxies                    
selected in the near-IR (e.g. in the H-band at 1.6 $\mu$m) with a                                 
limiting magnitude appropriate to cover the desired redshift range.                            
Examples of this approach are given by instruments where the efficient                         
multi-slit capability is provided by micro-shutter arrays (MSA) (JEDI;                         
Wang et al. 2004; Crotts et al. 2005; Cheng et al. 2006) or by 
digital micromirror devices (DMD) (SPACE; Cimatti et al. 2009). 
With the multi-slit approach, all galaxy types                           
(from passive ellipticals to starbursts) are observed, provided that the                       
targets are randomly selected from the magnitude-limited galaxy sample.                        
The second approach is based on slitless spectroscopy (e.g. Glazebrook et                      
al 2005; Laureijs et al. 2009; \citealt{JDEM})             
Due to stronger sky background, the slitless approach is                      
sensitive mostly to galaxies with emission lines (i.e. star-forming and                        
AGN systems), and uses mainly H$\alpha$ as a redshift tracer if the observations                    
are done in the near-IR to cover the redshift range of interest for dark                       
energy (e.g. $0.5<z<2$).                            

In this paper, we study the dark energy constraints
expected from plausible galaxy redshift surveys from space. We will
compare the various surveys using both the Dark Energy Task Force
(DETF) figure-of-merit (FoM) for ($w_0,w_a$) \citep{detf},
and more general dark energy FoMs motivated by the need to derive
model-independent constraints on dark energy \citep{Wang08a}.
In two accompanying papers, Majerotto et al. (2010) and 
\cite{Samushia10}, 
we will examine how space-based galaxy redshift surveys can test 
general relativity and are affected by cosmological model assumptions.

\section{Forecasting Methodology}

Galaxy redshift surveys allow us to determine the time dependence of
dark energy density by measuring the Hubble parameter $H(z)$ and the
angular diameter distance $D_A(z)=r(z)/(1+z)$ (where $r(z)$ is the
comoving distance) as a function of redshift based on Baryon Acoustic
Oscillation (BAO) measurements \citep{BG03,SE03}.  BAO in the observed
galaxy power spectrum provide a characteristic scale determined by the
comoving sound horizon at the drag epoch (shortly after
recombination), and are theoretically well understood. The signature
of the same physical process is clearly seen in the cosmic microwave
background (CMB) anisotropy data \citep{Komatsu10}, and these are
often used to anchor low-redshift BAO to the epoch of last
scattering. The observed radial BAO scale measures $sH(z)$ in the
radial direction, and $D_A(z)/s$ in the transverse direction, where
$s$ is the sound horizon at the baryon drag epoch. 
Redshift-space distortions (RSD) produced by linear peculiar velocities               
(Kaiser 1987) have also been shown in recent years to represent a                      
potentially powerful test of deviations from general relativity, the                   
alternative way to explain the observed cosmic acceleration (Guzzo et al.              
2008; Wang 2008b; Song \& Percival 2009; Reyes et al. 2010).  A large, deep             
redshift survey will be able to use RSD to measure the growth rate of                  
density fluctuations $f_g(z_i)$ within the same redshift bins in which                 
$H(z)$ will be estimated through BAO.           

The observed galaxy power spectrum can be reconstructed assuming a
reference cosmology, and can be approximated on
large scales as (see e.g. \citealt{SE03}):
\ba
\label{eq:P(k)b}
P_{obs}(k^{ref}_{\perp},k^{ref}_{\parallel}) &=&
\frac{\left[D_A(z)^{ref}\right]^2  H(z)}{\left[D_A(z)\right]^2 H(z)^{ref}}
\, b^2 \left( 1+\beta\, \mu^2 \right)^2
\cdot \nonumber\\
& \cdot& \left[ \frac{G(z)}{G(0)}\right]^2 P_{matter}(k)_{z=0}+ P_{shot},
\ea
where $b(z)$ is the bias factor between galaxy and matter
density distributions, and $\beta(z)$ is the linear redshift-space 
distortion parameter \citep{Kaiser}.
The growth factor $G(z)$ and the growth rate
$f_g(z)=\beta b(z)$ are related via
$f_g(z)=d\ln G(z)/d\ln a$, and 
$\mu = \bfk \cdot \hat{\bfr}/k$, with $\hat{\bfr}$ denoting the unit
vector along the line of sight; $\bfk$ is the wavevector with $|\bfk|=k$.
Hence $\mu^2=k^2_{\parallel}/k^2=k^2_{\parallel}/(k^2_{\perp}+k^2_{\parallel})$.
The values in the reference cosmology are denoted by the superscript ``ref'',
while those in the true cosmology have no subscript.
Note that 
\be
k^{ref}_{\perp}=k_\perp\,\frac{ D_A(z)}{D_A(z)^{ref}}, \hskip 0.5cm
k^{ref}_{\parallel}=k_\parallel\,\frac{H(z)^{ref}}{H(z)}.
\ee
The shot noise $P_{shot}$ is the unknown white shot noise that 
remains even after the conventional shot noise of inverse number density has been 
subtracted \citep{SE03}. These could arise from galaxy clustering bias even
on large scales due to local bias \citep{Seljak00}.
Eq.(\ref{eq:P(k)b}) characterizes the dependence of the observed galaxy power
spectrum on $H(z)$ and $D_A(z)$, as well as 
the sensitivity of a galaxy redshift survey to the redshift-space 
distortion parameter $\beta$.

The measurement of $f_g(z)$ given $\beta(z)$ requires an additional
measurement of the bias $b(z)$, which could be obtained from the
galaxy bispectrum \citep{Matarrese97,Verde02}. However, this masks the
fact that the redshift-space overdensity field has an additive
contribution that is independent of bias: galaxies move as test
particles in the matter flow, in a way that is independent of their
internal properties. The normalization of the redshift-space effect
depends on $f_g(z)\sigma_{8m}(z)$, and we rewrite Eq.(\ref{eq:P(k)b})
as
\ba
\label{eq:P(k)}
P_{obs}(k^{ref}_{\perp},k^{ref}_{\parallel}) &=&
\frac{\left[D_A(z)^{ref}\right]^2  H(z)}{\left[D_A(z)\right]^2 H(z)^{ref}}
C_0(k) \cdot \\
& \cdot& \left[ \sigma_{8g}(z)+ f_g(z)\sigma_{8m}(z)\, \mu^2 \right]^2 
 + P_{shot},\nonumber
\ea
where we have defined
\ba
&&\sigma^2_{8m}(z) \equiv \int_0^\infty \frac{\rd k}{k} \Delta^2(k|z)
\left[\frac{3 j_1(kr)}{kr}\right]^2,\\
&& \hskip 1cm r=8\,h^{-1}\mbox{Mpc},
\hskip 0.5cm \Delta^2(k|z) \equiv \frac{k^3 P_{m}(k|z)}{2\pi^2}.\\
&& C_0(k)\equiv \frac{P_{m}(k|z)}{\sigma^2_{8m}(z)}
=\frac{P_{m}(k|z=0)}{\sigma^2_{8m}(z=0)},
\ea
where $j_1(kr)$ is spherical Bessel function. Note that 
\be
\sigma_{8g}(z)=b(z) \sigma_{8m}(z).
\ee
We have assumed linear bias for simplicity.

To study the expected impact of future galaxy redshift surveys,
we use the Fisher matrix formalism.
In the limit where the length scale corresponding to
the survey volume is much larger than
the scale of any features in $P(k)$, 
we can assume that the likelihood function for the band powers of a 
galaxy redshift survey is Gaussian \citep{FKP}. 
Then the Fisher matrix can be approximated as \citep{Tegmark97}
\be
F_{ij}= \int_{k_{min}}^{k_{max}}
\frac{\partial\ln P(\bfk)}{\partial p_i}
\frac{\partial\ln P(\bfk)}{\partial p_j}\,
V_{eff}(\bfk)\, \frac{d \bfk^3}{2\, (2\pi)^3}
\label{eq:full Fisher}
\ee
where $p_i$ are the parameters to be estimated from data, and 
the derivatives are evaluated at parameter values of the
fiducial model. The effective volume of the survey
\ba
V_{eff}(k,\mu) &=&\int d\bfr^3 \left[ \frac{n(\bfr) P(k,\mu)}{ n(\bfr) P(k,\mu)+1}
\right]^2
\nonumber\\
&=&\left[ \frac{ n P(k,\mu)}{n P(k,\mu)+1} \right]^2 V_{survey},
\ea
where in the second equation, the comoving number density $n$ is 
assumed to only depend on the redshift for simplicity.
Note that the Fisher matrix $F_{ij}$ is the inverse of the covariance matrix
of the parameters $p_i$ if the $p_i$ are Gaussian distributed.
Eq.(\ref{eq:full Fisher}) propagates the measurement
error in $\ln P(\bfk)$ (which is proportional to $[V_{eff}(\bfk)]^{-1/2}$)
into measurement errors for the parameters $p_i$.

To minimize nonlinear effects, we restrict wavenumbers to the 
quasi-linear regime. We take $k_{min}=0$, and $k_{max}$ is given by requiring 
that the variance of matter fluctuations in a sphere of radius $R$, 
$\sigma^2(R)= 0.25$, for $R=\pi/(2k_{max})$. This gives $k_{max} 
\simeq 0.1\,h$Mpc$^{-1}$ at $z=0$, and $k_{max} \simeq 0.2\,h$Mpc$^{-1}$ 
at $z=1$, well within the quasi-linear regime.
In addition, we impose a uniform upper limit of $k_{max}\leq 0.2\,h$Mpc$^{-1}$
(i.e. $k_{max}=0.2\,h$Mpc$^{-1}$ at $z>1$), 
to ensure that we are only considering the conservative linear regime
essentially unaffected by nonlinear effects.

The observed galaxy power spectrum in a given redshift shell centered
at redshift $z_i$ can be described by a set of parameters, 
\{$H(z_i)$, $D_A(z_i)$, $f_g(z_i)\sigma_{8m}(z_i)$, $\sigma_{8g}(z_i)$, 
$P_{shot}^i$, $\omega_m$, $\omega_b$, $n_s$, $h$\}, where
$\omega_m=\Omega_m h^2 \propto \rho_m(z=0)$ (matter density today), 
$\omega_b=\Omega_b h^2 \propto \rho_b(z=0)$ (baryon density today),  
$n_s$ is the power-law index of the primordial matter power spectrum, 
and $h$ is the dimensionless Hubble constant. 
Note that $P_m(k)\propto k^{n_s} T^2(k)$, with the matter transfer function $T(k)$ 
only depending on $\omega_m$ and $\omega_b$ \citep{EisenHu98},\footnote{The effect 
of massive neutrinos will be considered elsewhere.} if $k$ were in units
of $1/$Mpc, and if the dark energy dependence of $T(k)$ can be neglected. 

We marginalize over \{$\sigma_{8g}(z_i)$, $P_{shot}^i$\} in each redshift
slice, and project \{$H(z_i)$, $D_A(z_i)$, $f_g(z_i)\sigma_{8m}(z_i)$, 
$\omega_m$, $\omega_b$, $n_s$, $h$\} into a final set of cosmological 
parameters \citep{Wang06}. We refer to this as the ``full $P(k)$ method,
with growth information included'', in which the growth information is
included assuming that general relativity is valid.
For more conservative dark energy constraints, we do not assume general relativity, 
and marginalize over \{$f_g(z_i)\sigma_{8m}(z_i)$\} from each redshift slice 
(in addition to \{$\sigma_{8g}(z_i)$, $P_{shot}^i$\}),
and only project \{$H(z_i)$, $D_A(z_i)$, $\omega_m$, $\omega_b$, $n_s$, $h$\} 
into the final set of cosmological parameters. 
We refer to this as the ``full $P(k)$ method, marginalized over growth information''.
The details of our implementation can be found in \cite{Wang06,Wang08a}. 
For an ultra conservative approach, we can marginalize over the 
cosmological parameters that describe the shape of the power spectrum,
\{$\omega_m$, $\omega_b$, $n_s$, $h$\}, and only project
\{$H(z_i)$, $D_A(z_i)$\} or \{$H(z_i)$, $D_A(z_i)$, $f_g(z_i)\sigma_{8m}(z_i)$\} 
into the final set of cosmological parameters. 
We refer to this as the ``$P(k)$-marginalized-over-shape" method.
To change from one set of parameters to another, we use \citep{Wang06}
\be
F_{\alpha \beta}= \sum_{ij} \frac{\partial p_i}{\partial q_{\alpha}}\,
F_{ij}\, \frac{\partial p_j}{\partial q_{\beta}}.
\label{eq:Fisherconv}
\ee
where $F_{\alpha \beta}$ is the Fisher matrix for a set of parameters $\bfp$, and 
$F_{ij}$ is the Fisher matrix for a set of equivalent parameters $\bfq$.

Measurements of the growth rate $f_g(z)$ and the BAO are correlated
and need to be considered simultaneously
\citep{ballinger96,Simpson10}. Note that the BAO only approach from
\cite{SE07} is similar to our ``$P(k)$-marginalize-over-shape'' approach, 
but we adopt a more general procedure that includes correlations between 
$\{H(z_i),D_A(z_i)\}$ and $\{f_g(z_i)\sigma_{8m}(z_i)\}$. Similarly, 
our approach is more general than that of 
White, Song \& Percival (2009), who made predictions for RSD constraints in a way 
that does not take into account simultaneous BAO measurements.                  

We derive dark energy constraints with and without Planck priors.
The Planck priors are included as discussed in Appendix \ref{sec:Planck}. 
We derive dark energy constraints with            
and without Planck priors, whose calculation is discussed in Appendix B. 
Given that Planck is already operating successfully, and the full Planck 
data will be available when a space-based galaxy survey is conducted 
(estimated to be around 2017), results including Planck priors are the 
most interesting for cosmological constraints. We have included results 
without Planck priors in order to show the level of the dependency on 
additional data, and to enable the reader to reproduce our results.

\section{Relative Importance of the Basic Survey Parameters}
\label{sec:basic}

Assuming the widely used linear dark energy equation of state
\citep{Chev01,Linder03},
\be
  w_X(z)=w_0+(1-a)w_a,
\label{eq:w0wa}
\ee
we now study the dependence of the DETF FoM for $(w_0,w_a)$ on the
basic survey parameters: redshift accuracy, minimum redshift of the
survey, and the survey area. We assume the fiducial cosmological model
adopted in the Euclid Assessment Study Report \citep{Laureijs09}:
$\Omega_m=0.25$, $\Omega_\Lambda=0.75$, $h=0.7$, $\sigma_8=0.80$,
$\Omega_b=0.0445$, $w_0=-0.95$, $w_a=0$, $n_s=1$.

We assume a baseline survey of H$\alpha$ emission line galaxies, based
on slitless spectroscopy of the sky. The empirical redshift
distribution of H$\alpha$ emission line galaxies derived by
\cite{Geach10} from observed H$\alpha$ luminosity functions was
adopted along with with the bias function derived by \cite{Orsi10}
using a galaxy formation simulation. 

Predictions for the redshift distribution of H$\alpha$ emitters are based on a simple model of      
the evolution of the observed H$\alpha$ luminosity function since $z\sim2$ (see Geach et al.\       
2010 for full details). Briefly, the model enforces a fixed space density over cosmic time, but     
allows $L^\star$ to increase with $(1+z)^Q$ evolution out to $z=1.3$ before plateauing at           
$z>1.3$. The exponent $Q$ is determined by fitting the evolution of observed $L^\star$ derived      
by different workers using similarly selected H$\alpha$ emitter samples over $0<z<1.3$. The         
1$\sigma$ uncertainty is determined by both the uncertainty of the observed $L^\star$ parameters    
and the redshift coverage windows of the various H$\alpha$ surveys employed. Combined with an       
uncertainty on the space density normalisation, we are able to estimate the typical error in        
${\rm d}N/{\rm d}z$ at a given limiting flux. Note that this does not include the uncertainty in    
the shape of the faint-end slope of the luminosity function, which is fixed at $\alpha=-1.35$ in    
the model. However, at the flux limits likely to be practical to future dark energy (galaxy         
redshift) surveys, galaxy counts contributed by $L<<L^\star$ galaxies will be negligible, and at    
$f_{\rm H\alpha}>10^{-16}$\,erg\,s$^{-1}$\,cm$^{-2}$ this simple model can successfully             
re-produce the observed number counts of H$\alpha$ emitters over the main redshift range            
pertinent to future dark energy (galaxy redshift) surveys.                                  

Orsi et al (2010) present predictions for the abundance and clustering of H-alpha                   
emitters using two different versions of their galaxy formation model. The two models contain many           
elements in common, but have important differences in their treatment of the formation of           
massive galaxies. One model invokes a superwind ejection of baryons to suppress the gas             
cooling rate in massive haloes, whereas the other model uses the energy released from               
accretion onto a central supermassive black hole. Orsi et al show that the predicted bias           
of H-alpha selected galaxies does not vary significantly between these models (the upper            
panels of their fig. 11) and is therefore a robust prediction.

Note that we consider the redshift success rates $e=0.35, 0.5, 0.7$ 
in all our results, thus effective varying the redshift completeness
over the entire plausible range. The uncertainties in the redshift distribution
and bias function of H$\alpha$ emission line galaxies are subdominant
compared to the uncertainty in the redshift success rate $e$, which
in turn depends on the mission implementation and survey strategy.

We present most of
our results in terms of the FoM for ($w_0,w_a)$, the conventional FoM
for comparing dark energy surveys proposed by the DETF
\citep{detf}. Fitting formulae are provided for $P(k)$ including
growth information (denoted ``FoM$_{P(k)f_g}$''), and when growth
information is marginalized over (``FoM$_{P(k)}$''). The effect of
extending the FoM definition is considered in Sec.\ref{sec:FoM_X}.

To include the ongoing Sloan Digital Sky Survey III (SDSS-III) Baryon
Oscillation Spectroscopic Survey (BOSS)\footnote{http://www.sdss3.org/cosmology.php} 
of luminous red galaxies (LRG) in our forecasts, we assume that the
LRG redshifts are measured over $0.1<z<0.7$, with
standard deviation $\sigma_z /(1+z)=0.001$, for a galaxy population
with a fixed number density of $n=3\times 10^{-4} h^3$Mpc$^{-3}$, and a
fixed linear bias of $b=1.7$, over a survey area of 10,000 (deg)$^2$.

\subsection{Dependence on area}

The FoM of $(w_0,w_a)$ for a survey is linearly dependent on the
effective survey volume $V_{eff}$ (see Eqs.[\ref{eq:full Fisher}] and
[\ref{eq:FoM}]), thus proportional to the survey area for a fixed
redshift range.
\begin{figure}
  \centering
  \includegraphics[trim = 0mm 0mm 60mm 0mm, width=0.9\columnwidth]{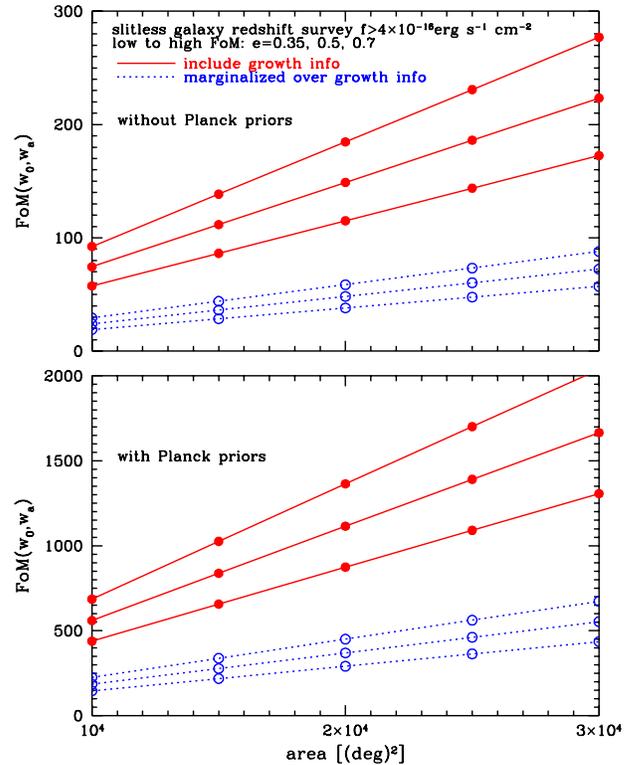}
  \caption{The FoM for $(w_0,w_a)$ for a slitless galaxy redshift
    survey as a function of the survey area.  We have assumed a survey
    of galaxies to a H$\alpha$ flux limit of 4$\times
    10^{-16}$erg$\,$s$^{-1}$cm$^{-2}$, with redshift success rates of
    $e=0.35, 0.5, 0.7$ to an accuracy of $\sigma_z/(1+z)=0.001$, over
    a redshift range of $0.5<z<2.1$.}
  \label{fig:FoM_w0wa_ENIS_planck_area}
\end{figure}
Fig.\ref{fig:FoM_w0wa_ENIS_planck_area} shows the FoM for $(w_0,w_a)$ for a 
slitless galaxy redshift survey as functions of the survey area.
We find that the dependence on survey area, with or without Planck priors,
is well approximated by
\be
\mbox{FoM} \propto \mbox[area].
\ee		

\subsection{Dependence on redshift accuracy}

\begin{figure*}
  \begin{center}
  \includegraphics[trim = 0mm 0mm 60mm 0mm, width=0.9\columnwidth]{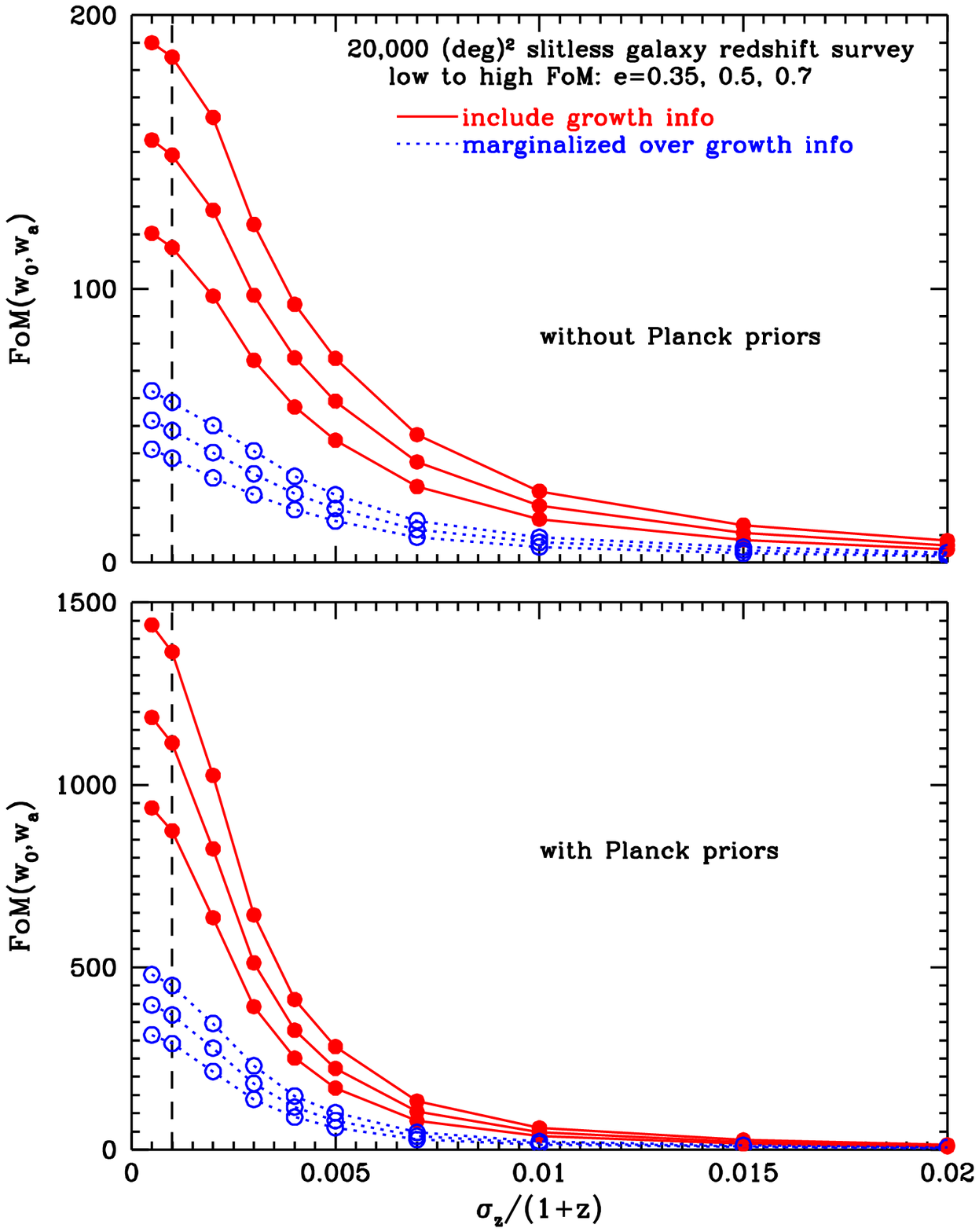}
  \hspace{1cm}
  \includegraphics[trim = 0mm 0mm 60mm 0mm, width=0.9\columnwidth]{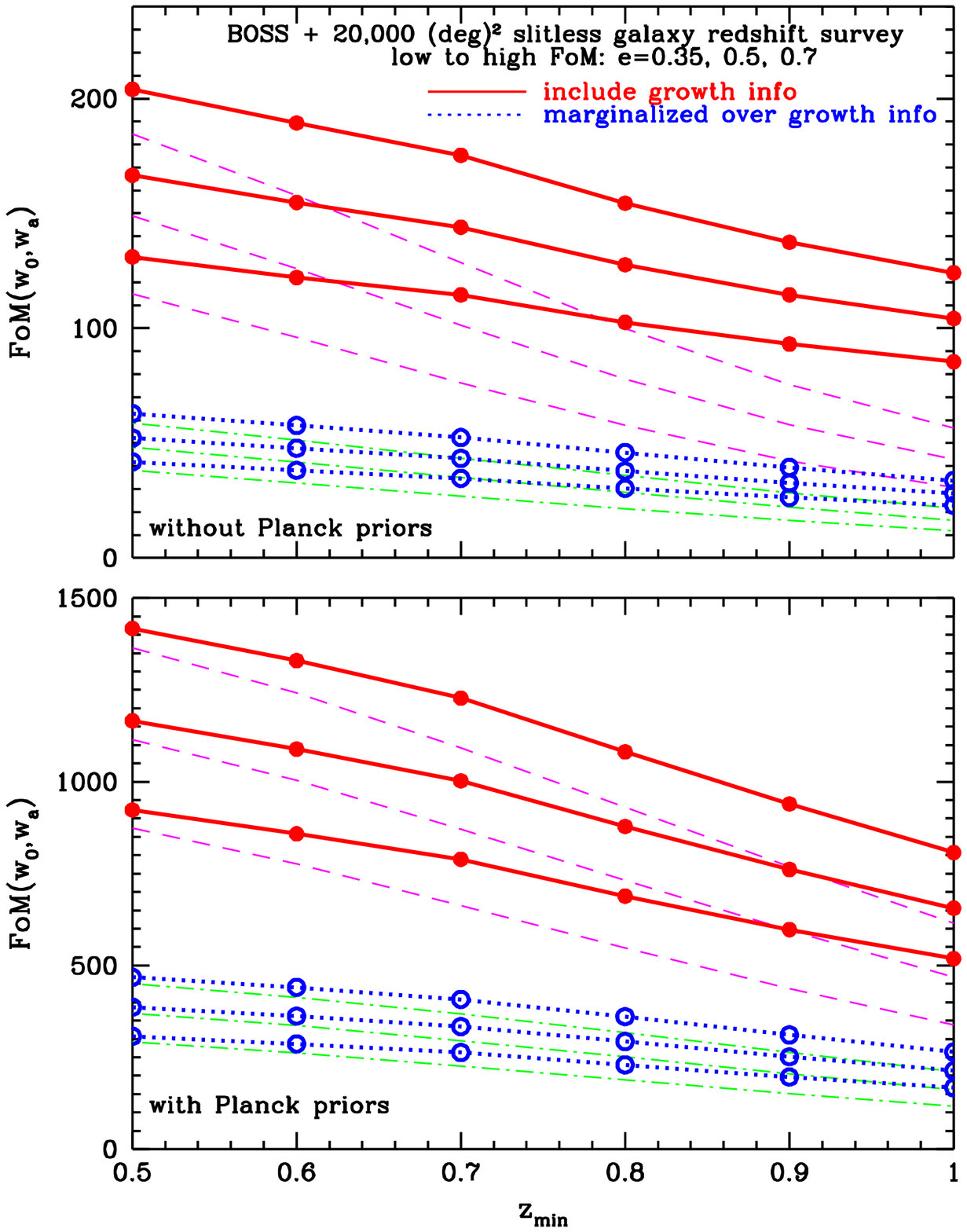}
  \end{center}
  \caption{The FoM for $(w_0,w_a)$ for a slitless galaxy redshift
    survey as functions of the redshift accuracy (left panels) and the
    minimum redshift (right panels) of the survey. We have assumed an
    H$\alpha$ flux limit of 4$\times
    10^{-16}$erg$\,$s$^{-1}$cm$^{-2}$, $z_{max}=2.1$, a survey area of
    20,000 (deg)$^2$, and redshift success rate $e=0.35, 0.5, 0.7$
    respectively. For the left panels, we have assumed $z_{min}=0.5$,
    and indicated our default assumption $\sigma_z/(1+z)=0.001$,
    assumed in the right panel, with vertical dashed lines.  The solid
    and dotted lines in each panel are the FoM for $(w_0,w_a)$ with
    growth information included and marginalized over respectively.
    Note that the right panels include BOSS data at $z\leq 0.5$; the
    slitless redshift survey only FoMs are represented by the dashed
    and dot-dashed curves.  The data points in the plots represent
    individual FoM calculations.}
\label{fig:FoM_w0wa_ENIS_planck_dlnz_zmin}
\end{figure*}
The left panels of Fig.\ref{fig:FoM_w0wa_ENIS_planck_dlnz_zmin} show the 
DETF FoM for $(w_0,w_a)$ for a slitless galaxy redshift survey with flux limit 
of 4$\times 10^{-16}$erg$\,$s$^{-1}$cm$^{-1}$, a survey area of 20,000
square degrees, and redshift success rates $e=0.35, 0.5, 0.7$
respectively, as functions of redshift accuracy (for $0.5\leq z \leq
2.1$). Appendix \ref{sec:fitting_FoM_dlnz} gives the fitting formulae for the
dependence of the FoM on the redshift accuracy for the various cases shown in 
the left panel of Fig.\ref{fig:FoM_w0wa_ENIS_planck_dlnz_zmin}.

The FoM increases rapidly as $\sigma_z$ decreases for
$\sigma_z/(1+z)\leq 0.001$, but the rate of increase slows down beyond
this limit (see left panel of
Fig.\ref{fig:FoM_w0wa_ENIS_planck_dlnz_zmin}). There is a minimum
redshift accuracy of $\sigma_z/(1+z)\simeq0.001$ that is is important
to achieve, but that further accuracy is not important if the cost is
high.

We have assumed that 35\%, 50\%, and 70\% (corresponding to
redshift success rates of $e$=0.35, 0.5, and 0.7) of objects have a 
correctly recovered redshift (with a redshift uncertainty $\sigma_z/(1+z)$), 
and that there is no contaminating fraction.                                   
Performance simulations made for the EUCLID mission show that 
redshift uncertainties are randomly distributed.                                                             
Given the objects' cross-contamination and the high background signal present                        
in slitless observation, the redshift measurement is much more difficult with                       
respect to the multi-slit case. To address this issue, a custom algorithm has                      
been developed by \cite{Franzetti10}, which is strongly linked to the detection 
of the H$\alpha$ line within the observational window. This algorithm 
selects high quality redshifts and makes 
line mis-identification very rare, and results in randomly distributed 
redshift failures \citep{Franzetti10}.

\subsection{Dependence on redshift range}

As discussed in Section~\ref{sec:intro}, one of the primary advantages
of a space-based survey is the ability to measure redshifts out to
$z\simeq2$. We therefore only consider changes to the minimum redshift
limit of the sample. Although there will always be a tail to low
redshift, we assume here that only redshifts greater than this minimum
are used to constrain DE models. The right panels in
Fig.\ref{fig:FoM_w0wa_ENIS_planck_dlnz_zmin} show the FoM for
$(w_0,w_a)$ as a function of the minimum redshift of galaxies within
the slitless survey assuming $z_{max}=2.1$.  The dashed and dot-dashed
curves are with growth information included and marginalized over
respectively. The solid and dotted curves are similar
to the dashed and dot-dashed curves, but include
BOSS data for $z \leq z_{min}$.  Appendix \ref{sec:fitting_FoM_zmin}
gives the fitting formulae for the dependence of the FoM on the
minimum redshift for the various cases shown in the right panel of
Fig.\ref{fig:FoM_w0wa_ENIS_planck_dlnz_zmin}.

The low redshift data have a strong
effect on the DETF FoM, and the inclusion of BOSS becomes increasingly
important as the minimum redshift of the slitless galaxy redshift
survey is increased beyond $z=0.7$, the maximum redshift covered by
BOSS. It is clear that, purely based on the DETF FoM, it would be
optimal to observe galaxies at lower redshifts. The bias of the DETF
FoM to low redshifts has been discussed many previous times
(e.g. \citealt{Albrecht09}), and ignores the power of a space-based
survey, as discussed in Section~\ref{sec:intro}. This is a situation
where it is obviously important to consider practical and instrumental
issues, as well as a comparison with what can be achieved from the
ground.

The redshift range of the survey of galaxies selected using a given
method is usually fixed and derived from instrumentation.  For
example, for H$\alpha$ flux selected galaxies observed from space, a
wavelength range between 1 and 2 $\mu$m driven by technical
considerations, naturally imposes a redshift range $0.52<z<2.05$ in
which H$\alpha$ will be visible \citep{Laureijs09}.  The
right panel of Fig.\ref{fig:FoM_w0wa_ENIS_planck_dlnz_zmin} shows
that, given this optimization method, it is better to choose the
smallest minimum redshift allowed by the instrumentation, even when it
overlaps with a ground-based survey.

Other arguments that should be considered for the optimal choice                              
of the redshift range are: (i) the capability to overlap (at least                             
partly) with other complementary surveys which sample galaxies with                            
a different biasing factor (e.g. BOSS/BigBOSS sampling Luminous Red                            
Galaxies at $0.1<z<1$ + EUCLID-like sampling star-forming galaxies at                            
$0.5<z<2$), (ii) the maximization of the redshift range in order to                              
have the largest leverage to constrain the potential evolution of                              
the dark energy density. The importance of these considerations
are demonstrated in Sec.\ref{sec:ground}.

\subsection{Dependence on flux limit}

We consider surveys with flux limits of 1, 2, 3, 4, and 5$\times
10^{-16}$erg$\,$s$^{-1}$cm$^{-2}$, and redshift success rates (the
percentages of galaxies for which the measured redshifts have the
specified redshift accuracy $\sigma_z$) of $e=0.35, 0.5, 0.7$.  Note
that for simplicity, we have assumed the redshift success rates are
constant with redshift.  Clearly, a realistic success rate in
measuring redshifts from a slitless survey will depend on
H$\alpha$ flux and redshifts. Ongoing simulations, however, show that
the overall trends and relative FoMs discussed here are not
significantly altered (Garilli et al., private                       
communication).

\begin{figure*}
  \begin{center}
  \includegraphics[trim = 0mm 0mm 60mm 0mm, width=0.9\columnwidth]{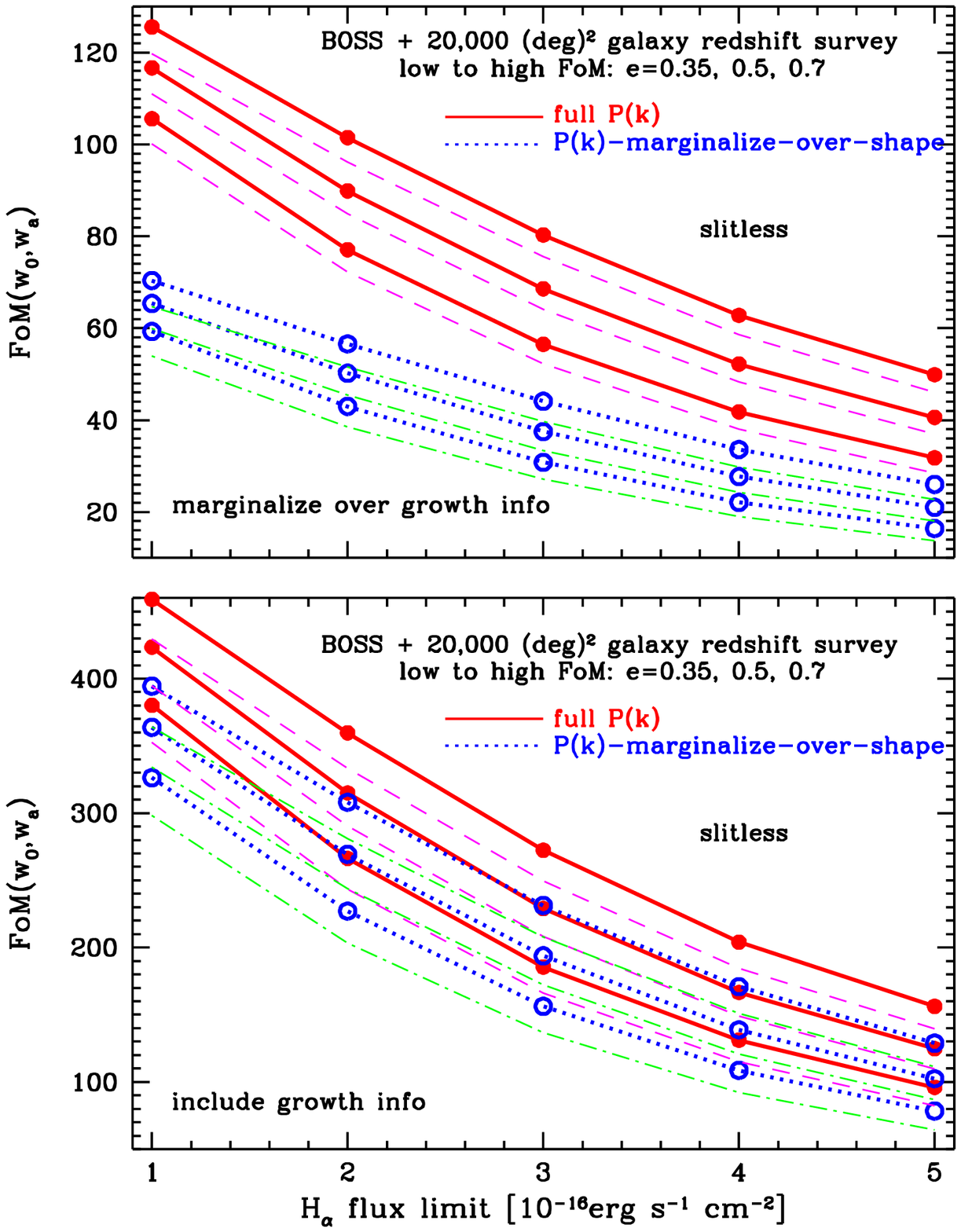}
  \hspace{1cm}
  \includegraphics[trim = 0mm 0mm 60mm 0mm, width=0.9\columnwidth]{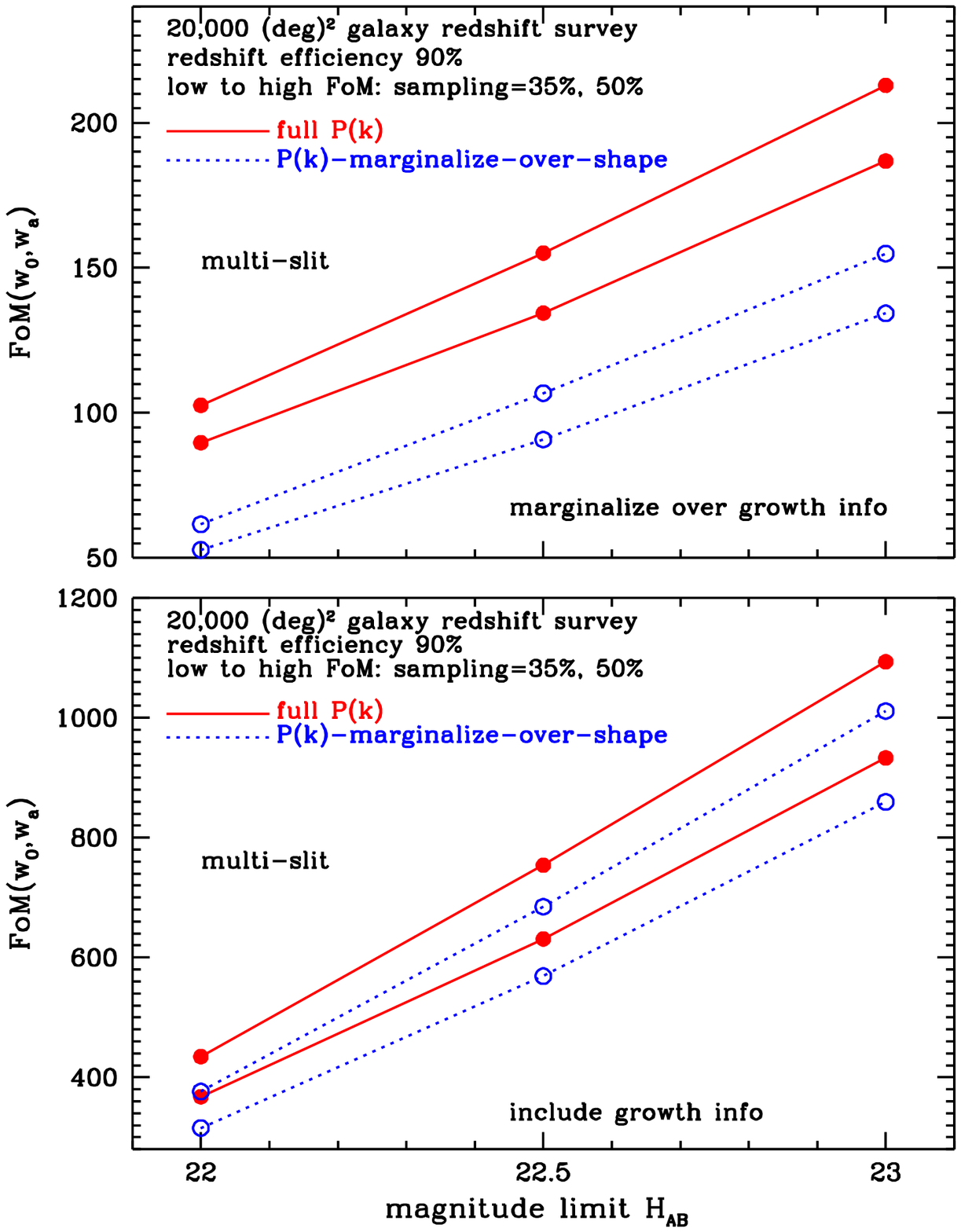}
  \end{center}
  \caption{The FoM for $(w_0,w_a)$ for a slitless galaxy redshift
    survey (left) as functions of the H$\alpha$ flux limit, and a multi-slit
    galaxy redshift survey (right) as functions of the H-band
    magnitude limit of the survey.  We have assumed
    $\sigma_z/(1+z)=0.001$, and a survey area of 20,000 (deg)$^2$. For
    the slitless survey, we have assumed $0.5<z<2.1$, redshift success
    rate $e=0.35, 0.5, 0.7$ respectively.  For the multi-slit survey, we have
    assumed a redshift efficiency of 90\%, and a redshift sampling
    rate of 35\% and 50\% respectively.  The solid and dotted lines in
    each panel are the FoM for $(w_0,w_a)$ with growth information
    included and marginalized over respectively.  Note that the left
    panel includes BOSS data at $z\leq 0.5$; the slitless redshift
    survey only FoMs are represented by the dashed and dot-dashed
    curves. Planck priors are not included.}
\label{fig:FoM_w0wa_ENIS_flux}
\end{figure*}
The left panel of Fig.\ref{fig:FoM_w0wa_ENIS_flux} shows the effect of
changing the H$\alpha$ flux limit to a slitless survey (without
adding Planck priors), assuming that
the data is taken to a uniform depth. 
Appendix \ref{sec:fitting_FoM_flux} gives the fitting formulae for
the dependence of the FoM on the H$\alpha$ flux limit for the various 
cases shown in the left panel of Fig.\ref{fig:FoM_w0wa_ENIS_flux}, with 
and without Planck priors.

As we saw previously for the minimum redshift, the addition of BOSS
data to a slitless galaxy redshift survey makes a notable improvement
on the FoM for ($w_0,w_a)$ covering the low redshift range where
$H(z)$ is more sensitive to dark energy if dark energy evolution is
small. Our fiducial model assumes a constant dark energy equation of
state $w=-0.95$, which implies a very weak evolution in the dark
energy density function $X(z)$.  

\subsection{Dependence on spectroscopic method}

We compare against a H-band magnitude limited survey of randomly
sampled galaxies enabled by multi-slit spectroscopy, e.g., by means of
programmable digital micromirror devices (DMD) (SPACE;
\citealt{Cimatti09}), or micro shutter arrays (MSA)
(JEDI; \citealt{Wang04,Crotts05,Cheng06}). To predict galaxy densities for such
surveys we use the empirical galaxy redshift distribution compiled by
Zamorani et al. from existing data (see \citealt{Laureijs09}), and we use
predictions of galaxy bias from galaxy formation simulations \citep{Orsi10}.
Our adopted H-band selected galaxy redshift distribution 
has been compiled from observations in the COSMOS survey and 
the Hubble Ultra-Deep Field, where excellent photometric redshifts are 
available.                                     
The bias function for H$\alpha$ flux and H-band magnitude selected
galaxies increase with redshift, with the former being less strongly
biased than the latter \citep{Orsi10}.  
The H-band traces massive structures (similar to selecting              
galaxies in the K-band), which makes them strongly biased. Star forming 
galaxies (which are selected by H$\alpha$ flux), on the other hand, appear 
to avoid the cores of clusters and populate the filaments of the dark matter 
structure, making them less biased than H-band galaxies \citep{Orsi10}. 
We consider multi-slit surveys with limiting magnitudes of $H_{AB}$=22,
22.5, and 23, a redshift success rate of 90\%, and sampling rates of
35\% and 50\%.

Fig.\ref{fig:FoM_w0wa_ENIS_flux} compares FoM for ($w_0,w_a)$ for
slitless and multi-slit galaxy redshift surveys (without
adding Planck priors). BOSS data are not
added to the multi-slit galaxy redshift surveys, which have redshift
ranges that extend to $z\sim 0.1$ \citep{Cimatti09,Laureijs09}.
Appendix \ref{sec:fitting_FoM_H} gives the fitting formulae for the
dependence of the H-band magnitude limit for the various cases shown
in the right panel of Fig.\ref{fig:FoM_w0wa_ENIS_flux},
with and without Planck priors.

The total number of galaxies with redshifts (measured with
$\sigma_z/(1+z)\le 0.001$) from a slitless survey is well
approximated by
\be
\frac{N_{gal}}{10^6}=276.74 \cdot\frac{[area]}{20000}\cdot
 \frac{e}{0.5} \cdot\left(\bar{f}\right)^{-0.9
 \left(\bar{f}\right)^{0.14}},
\ee
where $\bar{f}\equiv f/[10^{-16}$erg$\,$s$^{-1}$cm$^{-2}$].

The total number of galaxies with redshifts (measured with
$\sigma_z/(1+z)\le 0.001$) from a multi-slit survey is well
approximated by
\be
\frac{N_{gal}}{10^6}=\left[192.21+197.03\,(H_{AB}-22)^{1.3}\right]
\cdot\frac{[area]}{20000}\cdot\frac{e}{.9\times.35}.
\ee

Multi-slit surveys give significantly larger FoM than slitless surveys, 
because they allow the accurate redshift measurement for a greater number
of galaxies (and these galaxies are more biased tracers of large scale structure
than star-forming galaxies), and over a greater redshift range (extending
to $z\sim 0.1$). However, they have substantially stronger requirements in
instrumentation and mission implementation \citep{Cimatti09}.  

\subsection{Dependence on clustering information used}

\begin{figure}
  \centering
  \includegraphics[trim = 0mm 0mm 60mm 0mm, width=0.9\columnwidth]{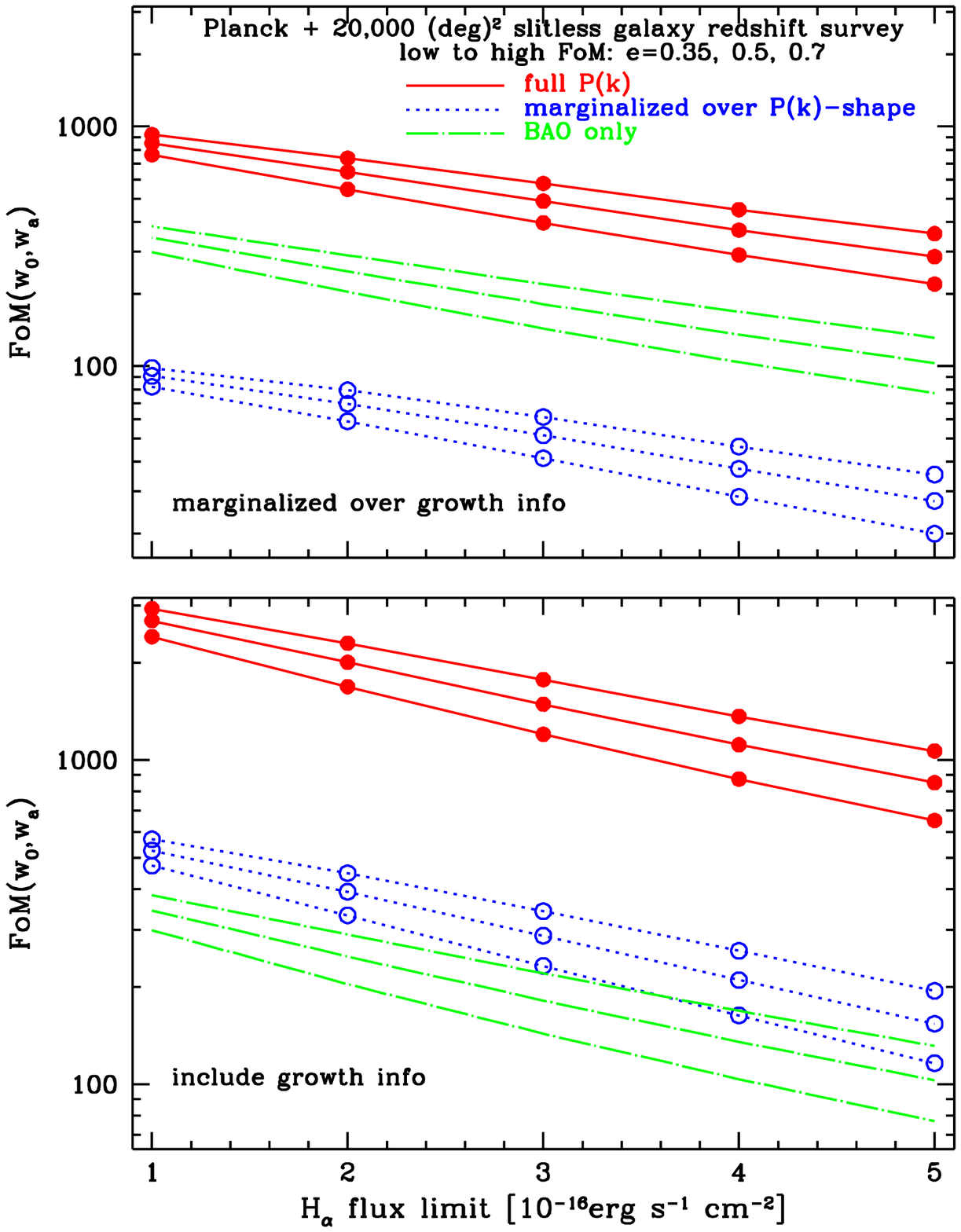}
  \caption{The FoM for $(w_0,w_a)$ for a slitless galaxy redshift
    survey combined with Planck priors, as functions of the H$\alpha$
    flux limit, for three different forecast methods.  We have assumed
    $0.5<z<2.1$, $\sigma_z/(1+z)=0.001$, a survey area of 20,000
    (deg)$^2$, and redshift success rate $e=0.35, 0.5, 0.7$ (curves from bottom to top)
    respectively.}
  \label{fig:FoM_w0wa_ENIS_planck}
\end{figure}
Fig.\ref{fig:FoM_w0wa_ENIS_planck} shows the FoM for $(w_0,w_a)$ for a
slitless galaxy redshift survey combined with Planck priors, as
functions of the H$\alpha$ flux limit, for three different levels of
clustering information used: the full $P(k)$ (solid lines),
$P(k)$-marginalized-over-shape (dotted lines), and BAO 
only (dot-dashed lines).  For the full $P(k)$ and
$P(k)$-marginalize-over-shape methods, the top panel of
Fig.\ref{fig:FoM_w0wa_ENIS_planck} shows the FoM obtained after
marginalization over growth information, while the lower panel shows
the FoM obtained including the growth information.  For the BAO only
method, the FoMs are the same in the upper and lower panels, and
obtained without adding the growth information, since the inclusion of
growth information is precluded by construction in this method: In the
BAO only method, the power spectrum with baryonic features is
approximated by \citep{SE07} \be
\label{eq:P_wg}
P_b(k,\mu|z) \propto \frac{\sin(x)}{x},
\hskip 1cm
x \equiv  \left(k^2_\perp s ^2_\perp +k^2_\parallel s^2_\parallel\right)^{1/2}.
\ee
The only parameters estimated in this method are the BAO scales in the
transverse and radial directions, $s_\perp$ and $s_\parallel$. To
include growth information, the Fisher matrix needs to be expanded to
include $f_g(z)\sigma_{8m}(z)$ and $\sigma_{8g}(z)$ for each redshift
slice.  However, RSD affect the amplitude of the full power spectrum,
without the damping factor of $\sin(x)/x$ in the BAO approximation of
Eq.~(\ref{eq:P_wg}). While this damping factor does not affect the
predictions of $s_\perp$ or $s_\parallel$ (the derivative of
$\sin(x)/x$ by $s_\perp$ or $s_\parallel$ is independent of $k$ to
leading order), it would incorrectly affect predictions of
$f_g(z)\sigma_{8m}(z)$ and $\sigma_{8g}(z)$ if this formulae was
naively applied to predict growth constraints. However, it is possible
to envisage a scenario where BAO are used to provide geometrical
constraints, while a coupled measurement of RSD is used based on the
full power spectrum.

When the growth information is marginalized over, the constraints from
the BAO only method are much stronger than those from the
$P(k)$-marginalize-over-shape method, with the addition of Planck
priors (see upper panel of Fig.\ref{fig:FoM_w0wa_ENIS_planck}).  This
is because the BAO only method implicitly assumes that the shape of
BAO (i.e., $P(k)$) are fixed by CMB data, while the
$P(k)$-marginalize-over-shape method allows the $P(k)$ shape to vary
and then discards the information of how cosmological constraints are
coupled to $P(k)$ shape. When growth information is included, the
information loss due to the marginalization over the shape of $P(k)$
is reduced, allowing a higher gain in FoM when Planck priors are added
(see lower panel of Fig.\ref{fig:FoM_w0wa_ENIS_planck}).

Fig.\ref{fig:FoM_w0wa_ENIS_planck} shows that both the
$P(k)$-marginalize-over-shape method and the BAO only
method give conservative estimates of dark energy
constraints. Compared to the BAO only method, the
$P(k)$-marginalize-over-shape method (including growth information)
has the advantage of allowing the consistent inclusion of growth
information in that, if we assume we can use the power spectrum shape
and amplitude to obtain RSD information, then it is sensible to also
assume we can at least partially use it as a standard ruler.

\subsection{Dependence on DE parametrization}
\label{sec:FoM_X}

We now consider the effect of changing to the generalized DE
parametrization \citep{Wang08a}, with the dimensionless dark energy
density $X(z)\equiv \rho_X(z)/\rho_X(z=0)$ given by interpolating its
value at $z_i$, $i=1$, 2, ..., $N$.  We consider $z_i=i\times 2.0/N$
($i=1, 2, ..., N$) with $N=3$, i.e., ($X_{0.67},X_{1.33},X_{2.0}$).
We use linear interpolation here since it gives the most conservative
estimates. Using this parametrization, we can define a dark energy
FoM \citep{Wang08a} \be \rmFoM(p_1,p_2,p_3,...)=\frac{1}{\sqrt{
    \rmdet\,\rmCov(p_1,p_2,p_3,...)}},
\label{eq:FoM}
\ee
where $\{p_i\}$ are the chosen set of dark energy parameters.
This definition has the advantage of being easy to calculate for
either real or simulated data, and applicable to any set of
dark energy parameters. 
If the likelihood surfaces for all the parameters are Gaussian,
this FoM is proportional to the inverse of the $N$-dimensional volume
enclosed by the 68\% confidence level (C.L.) contours of the parameters
$(p_1,p_2,p_3,...)$.
For $N=2$ and $(p_1,p_2)=(w_0,w_a)$, Eq.(\ref{eq:FoM}) reduces
to the FoM used by the DETF \citep{detf}, 
$\rmFoM(w_0,w_a)$.\footnote{The DETF defined the dark energy FoM
to be the inverse of the area enclosed by the 95\% C.L. contour of ($w_0,w_a$),
which is equal to the FoM given by Eq.(\ref{eq:FoM}) multiplied by a
constant factor of $1/(6.17\pi)$. However, this constant factor is
always omitted, even in the tables from the DETF report \citep{detf}.
Thus for all practical purposes, Eq.(\ref{eq:FoM}) is the same
as the DETF FoM for ($w_0,w_a$).}

\begin{figure*}
\begin{center}
  \includegraphics[trim = 0mm 0mm 60mm 0mm, width=0.9\columnwidth]{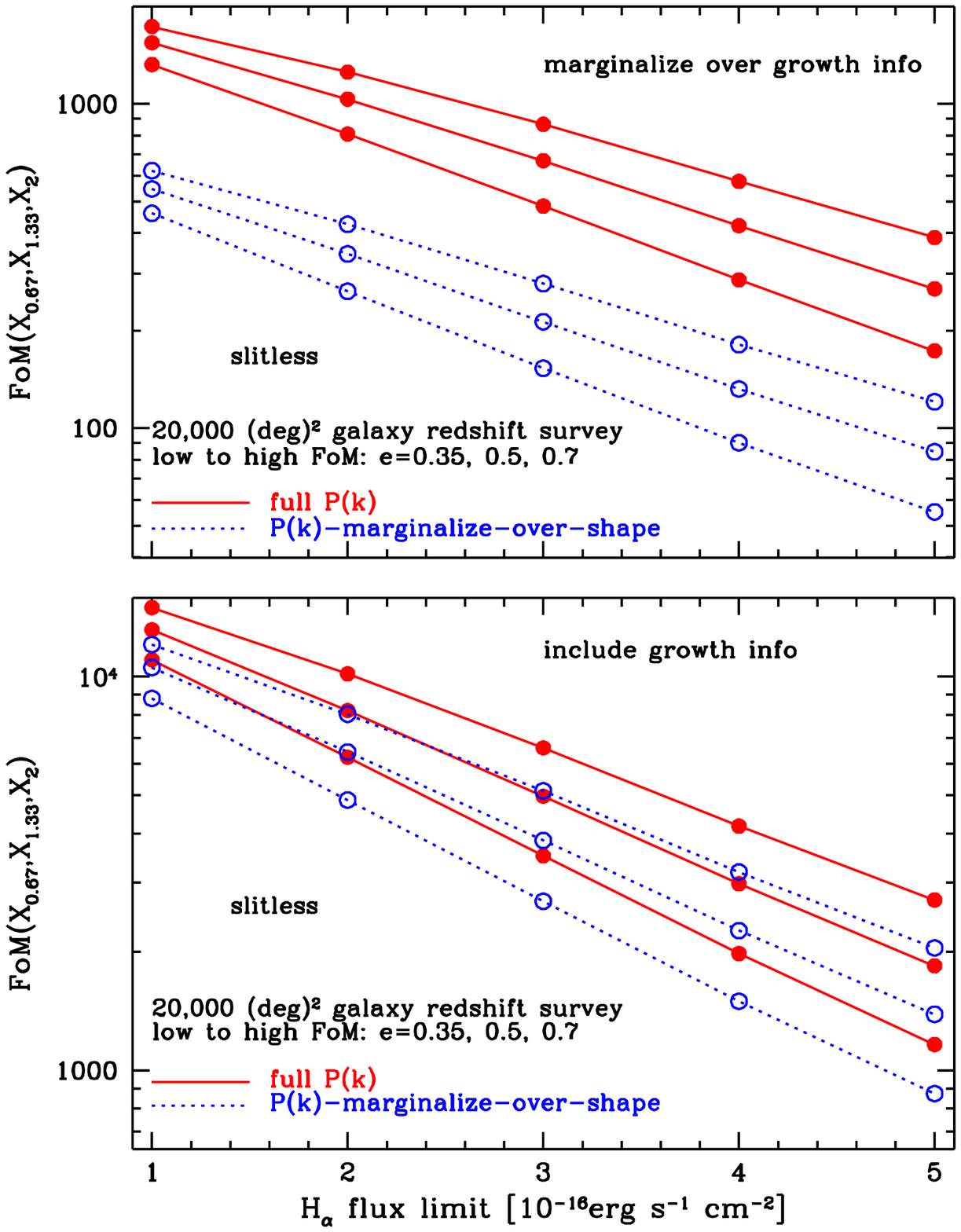}
  \hspace{0.5cm}
  \includegraphics[trim = 0mm 0mm 60mm 0mm, width=0.9\columnwidth]{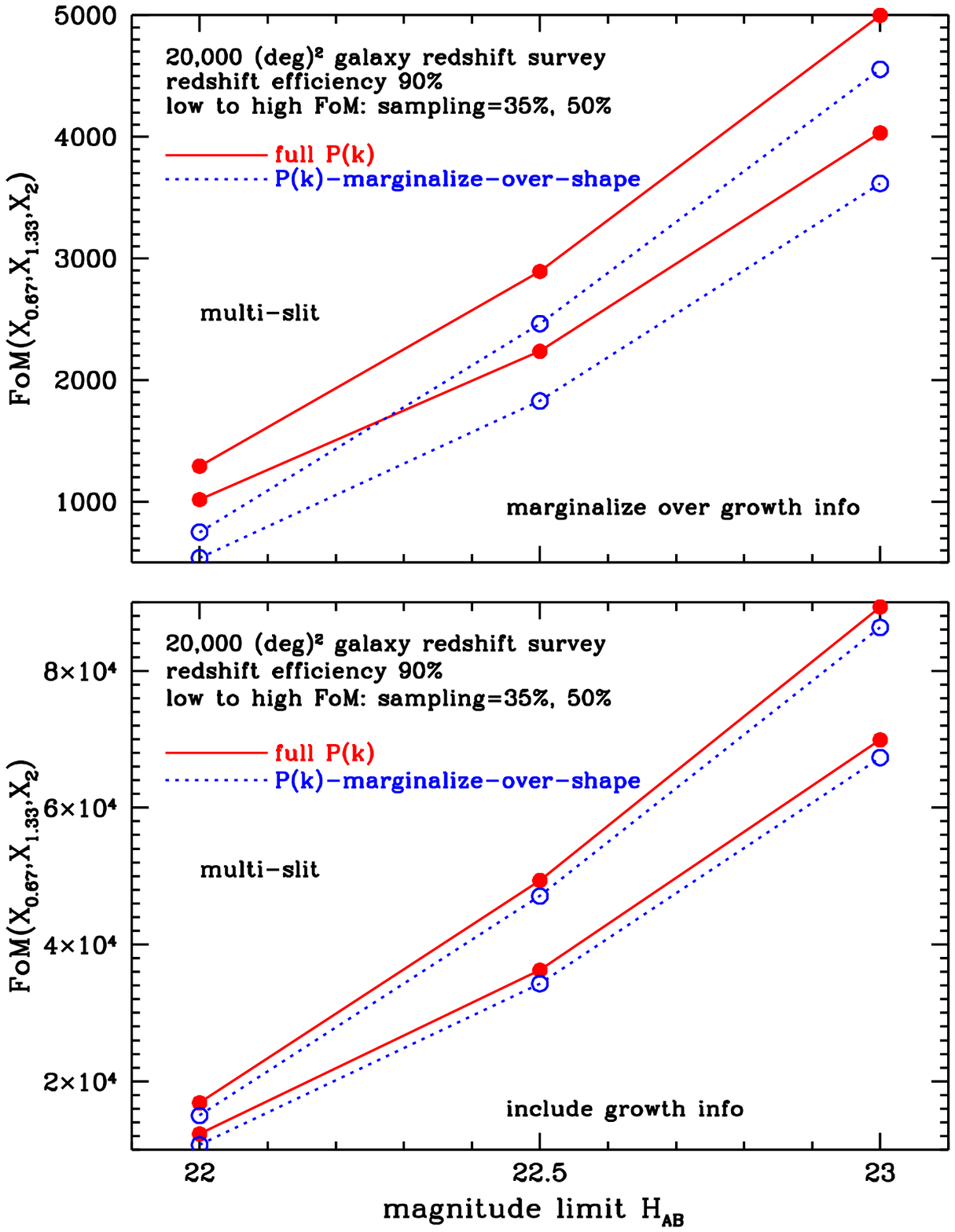}
\end{center}
\caption{The FoM for $\{X_{0.67},X_{1.33},X_{2.0} \}$ 
for a slitless galaxy redshift survey (left) as functions of the H$\alpha$ flux limit,
and for a multi-slit galaxy redshift survey (right) as functions of the H-band magnitude 
limit of the survey. We have assumed a survey area of 20,000 (deg)$^2$
and $\sigma_z/(1+z)=0.001$. For the slitless survey, we have assumed $0.5<z<2.1$, 
and redshift success rate $e=0.35, 0.5, 0.7$ respectively.
For the multi-slit survey, we have assumed a redshift efficiency of 90\%, and 
a redshift sampling rate of 35\% and 50\% respectively.}
\label{fig:FoM_X}
\end{figure*}
We now show the impact of parametrizing dark energy density function,
$X(z)\equiv \rho_X(z)/\rho_X(0)$, using its value at equally spaced redshifts,
($X(z_1),X(z_2),...,X(z_N)$).
Fig.\ref{fig:FoM_X} shows the FoM (without Planck priors)
for ($X_{0.67},X_{1.33},X_{2.0}$) 
for slitless (left panels) and multi-slit (right panels) galaxy 
redshift surveys, as functions of the H$\alpha$ flux limit for the slitless survey,
and of the H-band magnitude limit for the multi-slit survey. 
Note that the subscripts on $X$ indicate the redshift values.

We find that the addition of BOSS data to a slitless galaxy redshift survey
does {\it not} make a notable improvement on the FoM for ($X_{0.67},X_{1.33},X_{2.0}$), 
unlike the DETF FoM (see Fig.\ref{fig:FoM_w0wa_ENIS_flux}).  Adding
additional parameters to parametrize $X(z)$ gives qualitatively
similar results. This difference arises because fits to $w_0$ and
$w_a$ tend to give strongly correlated results, decreasing the
redshift-range over which we are sensitive to DE changes. Fits using
the parameter set ($X_{0.67},X_{1.33},X_{2.0}$) tend to be less
correlated, increasing the sensitivity to the behavior of dark energy
at the highest redshifts at which dark energy is important. Note that
multi-slit surveys would allow us to probe dark energy density at
$z>2$ by adding $X_{2.5}$ and $X_{3.0}$ to our parameter set.

The FoM for ($X_{0.67},X_{1.33},X_{2.0}$) is large, mainly because 
it involve an extra parameter compared to ($w_0,w_a$). For parameters 
$(p_1,p_2,p_3,...)$ that are well constrained by a survey, FoM$(p_1,p_2,p_3,...)$
roughly scales as $1/[\sigma(p_1)\sigma(p_2)\sigma(p_3)...]$. 

\subsection{Comparison with ground-based surveys}
\label{sec:ground}

Tables 1 and 2 compare the dark energy
constraints from fiducial space-based slitless and multi-slit surveys to that
of a generic ground-based survey. The space-based slitless and multi-slit surveys
are those considered by Euclid \citep{Laureijs09}. The fiducial slitless survey
is assumed to have an H$\alpha$ flux limit of 
4$\times 10^{-16}$erg$\,$s$^{-1}$cm$^{-2}$ (at 7$\sigma$), a redshift range of
$0.5<z<2.1$, $\sigma_z /(1+z)=0.001$, redshift efficiency $e=0.5$,
and a survey area of 20,000 (deg)$^2$.
The fiducial multi-slit survey is assumed to have an H-band magnitude limit 
of $H_{AB}=22$ (at 5$\sigma$), redshift success rate of 90\%, sampling rate of 35\%, 
$\sigma_z /(1+z)=0.001$, and a survey area of 20,000 (deg)$^2$.   

The generic ground-based survey has a redshift range of $0.1<z<1.4$,
$\sigma_z /(1+z)=0.001$, a fixed galaxy number density of $n=3\times
10^{-4} h^3$Mpc$^{-3}$, a fixed linear bias of $b=1.7$,\footnote{The
current best estimate for the low-$z$ SDSS LRGs is $b=1.7$ \citep{Reid10}.
Extending this assumption to higher redshift LRGs is conservative for 
passive evolution of LRGs. For emission line galaxies, however, 
$1 \la b \la 1.3$ for $0.7\leq z \leq 1.4$ \citep{Orsi10}.}
and a survey area of 10,000 (deg)$^2$.  Such a galaxy redshift survey can be
conducted using a single ground-based telescope, and select galaxies
based on standard optical colors\footnote{The proposed
PAU project will measure the redshifts of red, early-type galaxies 
in the interval $0.1 < z < 0.9$, with $\sigma_z/(1 + z) < 0.003$ (achieved
using 40 narrow filters and two broad filters),
and cover 8000 (deg)$^2$ \citep{Benitez09}.}. This essentially extends the BOSS
survey of LRGs from $z=0.7$ to $z=1.4$.  If more than a single telescope is
available for a ground-based survey, the survey area can be
significantly larger than 10,000 (deg)$^2$ (see, e.g.,
\citealt{Schlegel09}), then the FoM for dark energy will be increased by
the same factor as it scales with the survey area. While it is possible
for ground-based surveys to achieve similar area coverage as a
satellite mission, this would either require two instruments or a move
after completing one survey, causing the experiment to have a very
long duration.

\begin{table*}
\begin{center}
\begin{tabular}{crrrr|rrrr}
\hline
method & $dw_0$& $dw_a$ & $dw_p$ &FoM($w_0,w_a$)& $dw_0$ & $dw_a$ & $dw_p$ & FoM($w_0,w_a$)\\
\hline
&fiducial & slitless    & &&					+Planck&& & \\ 	 
$P(k)$ &      .103 &  .433 & .048  &  48.26  & .067  & .140  & .0193  & 369.58\\
$P(k)f_g$ &  .072 &  .274 & .024  & 148.93  & .023 &  .061 &  .0148 & 1114.91\\
\hline
&fiducial & multi-slit    & &&					+Planck&& & \\ 	 	 
$P(k)$ 	  &.078 & .318 & .035  &  89.70  &   .050  & .103 & .0169  & 576.44\\
$P(k)f_g$ & .049 & .182 & .015 &  367.51  &   .017 &  .044 & .0119 & 1907.60\\
\hline
&fiducial & ground    & &&					+Planck& && \\ 	 
$P(k)$   &   .182 &  .830 & .111 & 10.82  &    .156  & .360  & .0362   & 76.83\\
$P(k)f_g$ & .120  & .612 & .050 & 32.88   &   .049  & .130 &  .0305 &  253.28\\
\hline
\end{tabular}
\end{center}
\caption{The DETF FoM and 1-$\sigma$ marginalized errors for ($w_0,w_a$) for fiducial 
space-based slitless and multi-slit surveys, and a generic ground-based survey.}
\end{table*}
As a reference, the DETF found that the Stage II projects (current and ongoing surveys)
give FoM($w_0,w_a$) $\sim 50$ when combined with Planck priors \citep{detf}. 
Current data give FoM($w_0,w_a$) $\sim 10$-20 \citep{Wang09}.
Clearly, a space-based galaxy redshift survey (together with Planck data)
can potentially increase the FoM($w_0,w_a$) by a factor of $\sim$100 
compared to current data, and more than a factor of 10 compared to Stage II projects.

\begin{table*}
\begin{center}
\begin{tabular}{crrrr|rrrr}
\hline
method & $dX_{0.67}$ & $dX_{1.33}$ & $dX_{2.0}$ &FoM$\{X_i\}$& 
$dX_{0.67}$ & $dX_{1.33}$ & $dX_{2.0}$ & FoM$\{X_i\}$\\
\hline
&fiducial & slitless    & &&					+Planck& && \\ 	 
$P(k)$ &   .115  & .287 &  .624&  421.26 & .059 &   .058&    .163  &  3487.41\\   
$P(k)f_g$ & .055 & .164 &  .389 &2979.49 & .028  &  .046 &   .101&   26659.23 \\
\hline
&fiducial & multi-slit    & &&					+Planck& & &\\ 	 	 
$P(k)$ 	  &.080 &  .187  & .422&   1017.92 &  .052  & .051&  .125 &  6657.24\\
$P(k)f_g$ &.032 &  .082  & .201&  12300.00 &  .024  & .038 & .081 & 59674.89  \\
\hline
&fiducial & ground    & &&					+Planck&& & \\ 	 
$P(k)$   &  .232&   .600  &  6.011 &    3.46   & .106 &  .174  &  4.703   &   27.92 \\
$P(k)f_g$ & .132 &  .400 &   2.596&	31.14  & .049 &  .099 &   1.703 &    246.15\\
\hline
\end{tabular}
\end{center}
\caption{The FoM and 1-$\sigma$ marginalized errors for ($X_{0.67},X_{1.33},X_{2.0}$) 
for fiducial space-based slitless and multi-slit surveys, and a 
generic ground-based survey.}
\end{table*}

Note that the survey parameter assumptions adopted here
for space and ground based surveys do \emph{not}
reflect the limits of such surveys, but are just fiducial cases
generally considered feasible by the community.
We find that while a sufficiently wide ground-based survey 
(requiring more than one telescope) could give a similar FoM for ($w_0,w_a$) 
compared with a conservative space-based slitless survey, it will not
give competitive FoM for ($X_{0.67},X_{1.33},X_{2.0}$) (see Table 2). 

It is important to recognize that both space and ground galaxy
redshift surveys are required to obtain definitive measurement
of dark energy using galaxy clustering. 
Ongoing ground-based surveys, BOSS and WiggleZ\footnote{http://wigglez.swin.edu.au/}, 
will enable us to test the methodology for extracting dark energy constraints 
from galaxy clustering data, and improve our understanding of systematic effects.
Proposed ground-based surveys, such as BigBOSS \citep{Schlegel09} and
HETDEX\footnote{http://hetdex.org/}, will be complementary to space-based surveys
in using different tracer populations and redshift coverage. 

There are other tracers of cosmic large scale structure that can
be observed from the ground, and are also highly complementary in 
probing dark energy to the space-based surveys discussed in this paper.
For example, ground-based Ly$\alpha$ forest data can be used to study 
clustering of matter at $z= 2$ to 4 \citep{Croft02}, and help constrain the 
early evolution of dark energy.
Another example is the use of galaxy redshift surveys based on the radio 
HI emission line at 21 cm to probe dark energy. 
Galaxy Redshift Surveys made possible by the Square Kilometre 
Array (SKA) could use 21 cm emission to observe galaxies out               
to $z\sim 1.5$ \citep{Abdalla10}, but the time scale for such                     
experiments is longer than that of the currently proposed 
space-based surveys.             

The overlap in redshift ranges of space and ground-based surveys is
critical for understanding systematic effects such as bias
using multiple tracers of cosmic large scale structure
(e.g., H$\alpha$ selected galaxies from a space-based survey, and LRGs from
a ground-based survey). The use of multiple tracers of
cosmic large scale structure can ultimately increase the
precision of dark energy measurements from galaxy redshift surveys 
\citep{Seljak09}.

\section{Summary}

Recent studies (e.g. Cimatti et al. 2009) have shown that near-IR 
multi-slit spectroscopic surveys provide a very efficient               
approach for studying dark energy and would also provide data of 
sufficient quality for many other cosmological applications.                  
Slitless spectroscopy can also be very efficient and competitive if                            
some critical top level requirements are met such as the survey sky                            
coverage, the redshift accuracy, the number of galaxies.                                       
In particular, the combination of space-based survey and ground-based surveys
should encompass the entire redshift range, $0\la z \la 2$, in which 
dark energy becomes important. The ongoing ground-based survey
that covers the widest area (10,000 deg$^2$), BOSS, 
will span the redshift range of $0.1<z<0.7$.
The redshift range of $0.5\la z\la 2.1$ can be achieved by a space mission 
with near IR 1-2 $\mu$m wavelength coverage targeting H$\alpha$               
emission line galaxies \citep{Cimatti09}; expanding this redshift 
range would increase the complexity of a space mission. 

Our key findings from this paper are:
\begin{enumerate}

\item The redshift range of $0.5\la z \la 2.1$ is appropriate, 
since it exploits the redshift 
range that is only easily accessible from space, extends to sufficiently
low redshifts to allow both a vast 3-D map of the universe using a
single tracer population, and overlapping with ground-based
surveys such as BOSS and BigBOSS to enable reliable modeling of systematic
effects and increased statical precision. 

\item For a given survey depth, the dark energy FoM for $(w_0,w_a)$ increases 
linearly with the survey area. Thus it is optimal to cover the entire
extragalactic sky ($\sim$ 30,000 square degrees). 
The actual sky coverage of a given space-based survey will be 
constrained by cost and mission duration.
The sky coverage of $\geq$20,000 (degree)$^2$ can give powerful dark energy 
constraints (see Fig.\ref{fig:FoM_w0wa_ENIS_planck_area}).

\item There is a trade-off between survey area and survey depth
for a given mission implementation. Given the same total amount of exposure time,
maximizing the survey area gives the largest dark energy FoM, compared
to decreasing the survey area while increasing the survey depth.
The depth of the survey and the efficiency of the redshift measurements
are strongly constrained by the feasibility of the space mission 
instrumentation. We have assumed very conservative efficiencies 
for the redshift measurements \citep{Franzetti10}.
Taking into consideration the need to simplify the mission implementation
and reduce mission risks, a survey area of 20,000 (degree)$^2$
is feasible for a slitless survey with an H$\alpha$ flux limit of 
4$\times 10^{-16}$erg$\,$s$^{-1}$cm$^{-2}$, or a multi-slit survey
with an H-band magnitude limit of $H_{AB}<22$ \citep{Cimatti09,Laureijs09}.

\item A space-based galaxy redshift survey, optimized as discussed in
this paper, has enormous power in constraining dark energy 
(see Figs.\ref{fig:FoM_w0wa_ENIS_planck_dlnz_zmin}-\ref{fig:FoM_X} and Tables 1-2). 
The gain in dark energy FoM of an optimized space-based survey is most 
dramatic over ground-based surveys when dark energy density is allowed 
to be a free function parametrized by its values at equally spaced 
redshift values extending to $z=2$ (thus allowing a model-independent 
measurement of dark energy) (see Table 2).

\item The growth information from a galaxy redshift survey plays a critical
role in boosting the dark energy FoM of the survey, assuming that
general relativity is not modified. This is not surprising, since
existing measurements of the growth rate $f_g(z)$ have been used in
the past to help tighten dark energy constraints (see e.g.,
\citealt{Knop,WangPia04}). 
We find that when growth information from $P(k)$ is included in the analysis,
we gain a factor of $\sim$ 3 in the DETF dark energy FoM, compared
to when the growth information is marginalized over.
This is because the growth rate $f_g(z)$ is anti-correlated with 
$H(z)$. 

\item We show that in order to consistently include the
growth information, the full galaxy power spectrum $P(k)$ must
be used (i.e., the ``$P(k)$ method''). 
We can obtain conservative constraints if we marginalize 
over the cosmological parameters that determine the shape
of $P(k)$ (see Fig.\ref{fig:FoM_w0wa_ENIS_planck}).

\end{enumerate}

Probing dark energy using multiple techniques (galaxy clustering, weak lensing, 
supernovae), each with tight controls of systematic effects, will ultimately 
illuminate the nature of dark energy \citep{Wang04,Crotts05,Cheng06}.
A space-based galaxy redshift survey will play a key role in
advancing our understanding of dark energy.

\section*{Acknowledgements}

AC, CC, PF, BG, LG, EM, and GZ acknowledge the support from the Agenzia Spaziale Italiana                       
(ASI, contract N. I/058/08/0).   
CMB, JEG, CGL, and PM acknowledge support from the UK Science and Technology Facilities 
Research Council (STFC).
AO acknowledges a STFC Gemini studentship. 
PR acknowledges support by the DFG cluster of excellence Origin and Structure             
of the Universe.
LS is supported by European Research Grant,               
GNSF grant ST08/4-442 and SNSF SCOPES grant 128040. 
LS and WJP thank the European Research Council for financial support. WJP is
also grateful for support from the STFC and the Leverhulme Trust.

\setlength{\bibhang}{2.0em}

\appendix

\section{Fitting formulae for dark energy FoM}
\label{sec:fitting_FoM}

The fitting formulae for dark energy FoM correspond to the figures in
Sec.\ref{sec:basic}.

\subsection{Dependence of DETF FoM on the redshift accuracy}
\label{sec:fitting_FoM_dlnz}

The fitting formulae of the dependence of the DETF FoM on the redshift accuracy 
presented here correspond to the curves in left panels of 
Fig.\ref{fig:FoM_w0wa_ENIS_planck_dlnz_zmin}.
The dependence of the FoM for ($w_0,w_a$) on the redshift accuracy is well 
approximated by
\ba
&& \mbox{FoM}_{P(k)f_g} \nonumber\\
&=&\left\{ \begin{array}{ll}
 160.4 \left(\frac{e}{.5}\right)^{.64} 
\exp\left[-1.0453 \,x_z^{1.581(e/0.5)^{.177}}\right]&\\
 \hskip 0.5cm \mbox{for} \hskip 0.5cm 0.0005 \leq \sigma_z/(1+z) <0.005; &\\
  58.99\left(\frac{e}{.5}\right)^{.75}  x_z^{-1.6}, &\\
 \hskip 0.5cm \mbox{for} \hskip 0.5cm 0.005\leq \sigma_z/(1+z)\leq 0.02 & 
\end{array}\right.
\ea
\ba
&&\mbox{FoM}_{P(k)}\nonumber\\
&=&\left\{ \begin{array}{ll}
54.64 \left(\frac{e}{.5}\right)^{.55} 
\exp\left[-1.024 \,x_z^{1.313(e/.5)^{.247}}\right]&\\
 \hskip 0.5cm \mbox{for} \hskip 0.5cm 0.0005 \leq \sigma_z/(1+z) <0.005; &\\
19.78\left(\frac{e}{.5}\right)^{.73} x_z^{-1.366}, &\\
 \hskip 0.5cm \mbox{for} \hskip 0.5cm 0.005\leq \sigma_z/(1+z)\leq 0.02 &
\end{array}\right.
\ea	
where we have defined
\be
x_z \equiv \frac{\sigma_z/(1+z)}{0.005}.
\ee
When Planck priors are added (see Appendix \ref{sec:Planck}), we find
\ba
&&\mbox{FoM}_{P(k)f_g}\nonumber\\
&=&\left\{ \begin{array}{ll}
1288 \left(\frac{e}{.5}\right)^{.61} 
\exp\left[-1.815 \left(\frac{.5}{e}\right)^{.079}
x_z^{1.464(e/.5)^{.06}}\right]&\\
\hskip 0.5cm \mbox{for} \hskip 0.5cm 0.0005 \leq \sigma_z/(1+z) <0.005; &\\
222.91\left(\frac{e}{.5}\right)^{.75} x_z^{-2.165}, &\\
\hskip 0.5cm \mbox{for} \hskip 0.5cm 0.005\leq \sigma_z/(1+z) \leq 0.02 &
\end{array}\right.
\ea	
\ba
&&\mbox{FoM}_{P(k)}\nonumber\\
&=&\left\{ \begin{array}{ll}
428.7\left(\frac{e}{.5}\right)^{.59} 
\exp\left[-1.711 \left(\frac{.5}{e}\right)^{.103}
x_z^{1.443(e/.5)^{.076}}\right]&\\
\hskip 1cm \mbox{for} \hskip 0.5cm 0.0005 \leq \sigma_z/(1+z) <0.005; &\\
 79.9\left(\frac{e}{.5}\right)^{.73} x_z^{-1.92}&\\
\hskip 1cm \mbox{for} \hskip 0.5cm 0.005\leq \sigma_z/(1+z) \leq 0.02 &
\end{array}\right.
\ea	

\subsection{Dependence on minimum redshift}
\label{sec:fitting_FoM_zmin}

The fitting formulae of the dependence of the DETF FoM on the minimum redshift  
presented here correspond to the curves in right panels of 
Fig.\ref{fig:FoM_w0wa_ENIS_planck_dlnz_zmin}.
For a slitless galaxy redshift survey only,
the dependence of the FoM for ($w_0,w_a$) on $z_{min}$ is well approximated by
\ba
\mbox{FoM}_{P(k)f_g}&=&148.93 \left(\frac{e}{.5}\right)^{.68}
 -237.23 \left(\frac{e}{.5}\right)^{.56}\Delta z_{min}\cdot\nonumber\\
& &\hskip 0.5cm
\cdot\exp\left[-.12\left(\frac{\Delta z_{min}}{.5}\right)^3\right],\\
\mbox{FoM}_{P(k)}&=&48.26 \left(\frac{e}{.5}\right)^{.62}
-66 \left(\frac{e}{.5}\right)^{.56}\Delta z_{min}\cdot\,\,\,\,\nonumber\\
& &\hskip 0.5cm
\exp\left[-.05\left(\frac{\Delta z_{min}}{.5}\right)^3\right],
\ea
where we have defined
\be
\Delta z_{min} \equiv z_{min}-0.5.
\ee
When BOSS data are added, we find
\ba
\mbox{FoM}_{P(k)f_g}&=&166.62 \left(\frac{e}{.5}\right)^{.64}
-134 \left(\frac{e}{.5}\right)^{.84}\Delta z_{min},\\
\mbox{FoM}_{P(k)}&=&52.22 \left(\frac{e}{.5}\right)^{.59}
-56.1 \left(\frac{e}{.5}\right)^{.62}\Delta z_{min}\cdot\nonumber\\
& &\hskip 0.5cm
\exp\left[-.12\left(\frac{\Delta z_{min}}{.5}\right)^3\right].
\ea
For a slitless galaxy redshift survey with Planck priors 
(see Appendix \ref{sec:Planck}), we find
\ba
\mbox{FoM}_{P(k)f_g}=1114.91 \left(\frac{e}{.5}\right)^{.64}
 -1292.64 \left(\frac{e}{.5}\right)^{.48}\Delta z_{min}&\\
\mbox{FoM}_{P(k)}= 369.58 \left(\frac{e}{.5}\right)^{.63}
-415 \left(\frac{e}{.5}\right)^{.46}\Delta z_{min}.&
\ea
For a slitless galaxy redshift survey combined with BOSS data
and Planck priors (see Appendix \ref{sec:Planck} ), we find
\be
\mbox{FoM}_{P(k)f_g}=\left\{ \begin{array}{ll}
1165.83 \left(\frac{e}{.5}\right)^{.62}
-817.85\left(\frac{e}{.5}\right)^{.50}\Delta z_{min}&\\
\hskip 0.5cm \mbox{for} \hskip 0.5cm 0.5 \leq z_{min} \leq 0.7&\\
1002.26 \left(\frac{e}{.5}\right)^{.64}
-1155.23\left(\frac{e}{.5}\right)^{.82}( z_{min}-.7),&\\
\hskip 0.5cm \mbox{for} \hskip 0.5cm 0.7 < z_{min} \leq 1&
\end{array}\right.
\ee
\be
\mbox{FoM}_{P(k)}=\left\{ \begin{array}{ll}
386.53 \left(\frac{e}{.5}\right)^{.61}
-264.35 \left(\frac{e}{.5}\right)^{.44}\Delta z_{min}&\\
 \hskip 0.5cm \mbox{for} \hskip 0.5cm 0.5 \leq z_{min} \leq 0.7&\\
333.66 \left(\frac{e}{.5}\right)^{.63}
-400.27\left(\frac{e}{.5}\right)^{.58}( z_{min}-.7)&\\
\hskip 0.5cm \mbox{for} \hskip 0.5cm 0.7 < z_{min} \leq 1&
\end{array}\right.
\ee

\subsection{Dependence on the H$\alpha$ flux limit}
\label{sec:fitting_FoM_flux}

The fitting formulae of the dependence of the DETF FoM on the H$\alpha$
flux limit presented here correspond to the curves in left panels of 
Fig.\ref{fig:FoM_w0wa_ENIS_flux}.
For slitless galaxy redshift surveys, the dependence of the FoM for
($w_0,w_a$) on the H$\alpha$ flux limit is well approximated by
\ba
\mbox{FoM}_{P(k)f_g}=148.9 \left(\frac{e}{.5}\right)^{.68}
\exp\left[-.321\left(\frac{.5}{e}\right)^{.37}(\bar{f}-4)\right]
&\\
\mbox{FoM}_{P(k)}=48.3\left(\frac{e}{.5}\right)^{.62}
\exp\left[-.275\left(\frac{.5}{e}\right)^{.4}(\bar{f}-4)\right]&
\ea
where $\bar{f}\equiv f/[10^{-16}$erg$\,$s$^{-1}$cm$^{-2}$].  When
Planck priors (see Appendix \ref{sec:Planck}) are added to slitless galaxy redshift
surveys (not shown in Fig.\ref{fig:FoM_w0wa_ENIS_flux}), we find
\ba
\mbox{FoM}_{P(k)f_g}
=1114.9 \left(\frac{e}{.5}\right)^{.64}
\exp\left[-.288\left(\frac{.5}{e}\right)^{.37}(\bar{f}-4)\right]&\\
\mbox{FoM}_{P(k)}
=369.6\left(\frac{e}{.5}\right)^{.63}
\exp\left[-.273\left(\frac{.5}{e}\right)^{.39}(\bar{f}-4)\right]&.
\ea

\subsection{Dependence on H-band magnitude limit}
\label{sec:fitting_FoM_H}

The fitting formulae of the dependence of the DETF FoM on the 
H-band magnitude limit presented here correspond to the curves in the
right panels of Fig.\ref{fig:FoM_w0wa_ENIS_flux}.
For multi-slit galaxy redshift surveys, the dependence of the FoM for 
($w_0,w_a$) on the H-band magnitude limit can be approximated by
\ba
&&\mbox{FoM}_{P(k)f_g}\nonumber\\
&=&367.5 \left(\frac{e}{.315}\right)^{.47}
+565.6 \left(\frac{e}{.315}\right)^{.43}(H_{AB}-22) \\
&&\mbox{FoM}_{P(k)}\nonumber\\
&=&89.7 \left(\frac{e}{.315}\right)^{.38}
+97.1 \left(\frac{e}{.315}\right)^{.36}(H_{AB}-22).
\ea
Note that for multi-slit surveys, $e$ is the product of the
redshift efficiency and the redshift sampling rate.
Thus for a redshift efficiency of 90\%, and a redshift sampling rate
of 35\%, $e=0.315$.

When Planck priors (see Appendix \ref{sec:Planck}) are added to 
multi-slit galaxy redshift surveys (not shown in 
Fig.\ref{fig:FoM_w0wa_ENIS_flux}), we find 
\ba
&&\mbox{FoM}_{P(k)f_g}\nonumber\\
&=&1907.6 \left(\frac{e}{.315}\right)^{.4}
\exp\left[.71 \left(\frac{e}{.315}\right)^{.23}(H_{AB}-22) \right]\\
&&\mbox{FoM}_{P(k)}\nonumber\\
&=&576.4 \left(\frac{e}{.315}\right)^{.37}
\exp\left[.64 \left(\frac{e}{.315}\right)^{.34}(H_{AB}-22) \right]
\ea

\section{Planck priors}
\label{sec:Planck}

We derive and include Planck priors as discussed in \cite{Mukherjee08}.
Our simulation and treatment of Planck data is as in \cite{Pahud06}. 
We include the temperature and 
polarization (TT, TE, and EE) spectra from three temperature channels with 
specification similar to the HFI channels of frequency 100 GHz, 143 GHz, 
and 217 GHz, and one 143 GHz polarization channel, following the current
Planck documentation,\footnote{
www.rssd.esa.int/index.php?project=PLANCK\&page=perf\underline{~}top}.
The full likelihood is constructed assuming a sky coverage of 0.8. 
We choose a fiducial model to be the Euclid fiducial model:
$\Omega_m=0.25$, $\Omega_\Lambda=0.75$, $h=0.7$, 
$\sigma_8=0.80$, $\Omega_b=0.0445$, $w_0=-0.95$, $w_a=0$, $n_s=1$.

CMB data can be effectively and simply summarized in the context of
combining with other data by adding a term in the likelihood 
that involve the CMB shift parameter $R$, the angular scale of the 
sound horizon at last scattering $l_a$, the baryon density 
$\Omega_bh^2$, and the power-law index of the primordial matter
power spectrum $n_s$ \citep{WangPia06,WangPia07}. 
This method is independent of the dark energy model 
used as long as only background quantities are varied.
The CMB shift parameter $R$ is defined as
\be
R \equiv \sqrt{\Omega_m H_0^2} \,r(z_{CMB}), \hskip 0.1in
l_a \equiv \pi r(z_{CMB})/r_s(z_{CMB}),
\label{eq1}
\ee
where $r(z)$ is the comoving distance from the observer to redshift $z$,
and $r_s(z_{CMB})$ is the comoving size of the sound-horizon at decoupling.

The comoving distance to a redshift $z$ is given by
\ba
\label{eq:rz}
&&r(z)=cH_0^{-1}\, |\Omega_k|^{-1/2} {\rm sinn}[|\Omega_k|^{1/2}\, \Gamma(z)]\\
&&\Gamma(z)=\int_0^z\frac{dz'}{E(z')}, \hskip 1cm E(z)=H(z)/H_0 \nonumber
\ea
where $\Omega_k=-k/H_0^2$ with $k$ denoting the curvature constant, 
and ${\rm sinn}(x)=\sin(x)$, $x$, $\sinh(x)$ for 
$\Omega_k<0$, $\Omega_k=0$, and $\Omega_k>0$ respectively, and
\be
E(z)
=\left[\Omega_m (1+z)^3+\Omega_{\rm rad}(1+z)^4 
          +\Omega_k(1+z)^2+\Omega_X X(z)\right]^{1/2}
\ee
with $\Omega_X=1-\Omega_m-\Omega_{\rm rad}-\Omega_k$, and the dark energy density
function $X(z) \equiv \rho_X(z)/\rho_X(0)$.

We calculate the distance to decoupling, $z_{CMB}$, via the fitting 
formula in \cite{HuSugiyama} (same as that used by CAMB \citealt{camb}).
The comoving sound horizon at recombination is given by
\ba
\label{eq:rs}
r_s(z_{CMB})& =& \int_0^{t_{CMB}} \frac{c_s\, dt}{a}
=cH_0^{-1}\int_{z_{CMB}}^{\infty} dz\,
\frac{c_s}{E(z)} \nonumber\\
&=& cH_0^{-1} \int_0^{a_{CMB}} 
\frac{da}{\sqrt{ 3(1+ \overline{R_b}\,a)\, a^4 E^2(z)}},
\ea
where $a$ is the cosmic scale factor, 
$a_{CMB} =1/(1+z_{CMB})$, and
$a^4 E^2(z)=\Omega_{\rm rad}+ \Omega_m a+\Omega_k a^2 +\Omega_X X(z) a^4$.
The radiation density is computed using the Stefan-Boltzmann formula from the
CMB temperature, assuming $3.04$ families of massless neutrinos.
The sound speed is $c_s=1/\sqrt{3(1+\overline{R_b}\,a)}$,
with $\overline{R_b}\,a=3\rho_b/(4\rho_\gamma)$,
$\overline{R_b}=31500\Omega_bh^2(T_{CMB}/2.7\,{\rm K})^{-4}$.

We derived the full covariance matrix of $(R, l_a, \Omega_b h^2, n_s)$
through an MCMC based likelihood analysis \citep{LewisBridle} of 
simulated Planck data. 
The Fisher matrix of $\bfq=(R$, $l_a$, $\omega_b$, $n_s$) is the inverse of
the covariance matrix of $\bfq$.
Note that the CMB shift parameters $R$ and $l_a$ encode all the information on dark energy
parameters. For any given dark energy model parameterized by the parameter
set $\bfp_X$, the relevant Fisher matrix for $\bfp=(\bfp_X$, $\Omega_X$, $\Omega_k$, 
$\omega_m$, $\omega_b$, $n_S$) can be found using Eq.(\ref{eq:Fisherconv}).
Eq.(\ref{eq:w0wa}) and Sec.\ref{sec:FoM_X} describe our dark energy parametrization.

\label{lastpage}

\end{document}